\documentclass[onecolumn,notitlepage,showpacs,pre,preprintnumbers,amsmath,amssymb,aps,floatfix]{revtex4-1}

\pdfoutput=1
\usepackage{graphicx}
\usepackage{dcolumn}
\usepackage{bm}
\usepackage{array}
\usepackage{multirow}
\usepackage{graphicx,capt-of}
\usepackage{caption}
\usepackage{subcaption}
\usepackage{mwe} 
\usepackage{graphicx}\newcounter{subfloat}

\usepackage[T1]{fontenc}
\usepackage[font=small,labelfont=bf,tableposition=top]{caption}
\begin{document}
\newcommand\MyBox[2]{
  \fbox{\lower0.75cm
    \vbox to 1.7cm{\vfil
      \hbox to 1.7cm{\hfil\parbox{1.4cm}{#1\\#2}\hfil}
      \vfil}%
  }%
}
\newcommand{\Vh}[1]{\hat{#1}}
\newcommand{\Aa}{A^1_{\epsilon}}
\newcommand{\Ab}{A^L_{\epsilon}}
\newcommand{\Ae}{A_{\epsilon}}
\newcommand{\finn}[1]{\phi^{\pm}_{#1}}
\newcommand{\ea}{e^{-|\alpha|^2}}
\newcommand{\eb}{\frac{e^{-|\alpha|^2} |\alpha|^{2 n}}{n!}}
\newcommand{\ebbb}{\frac{e^{-|\alpha|^2} |\alpha|^{2 (l+n+m)}}{l!m!n!}}
\newcommand{\ass}{\alpha}
\newcommand{\mcP}{\mathcal{P}}
\newcommand{\mcC}{\mathcal{C}}
\newcommand{\as}{\alpha^*}
\newcommand{\fb}{\bar{f}}
\newcommand{\gb}{\bar{g}}
\newcommand{\la}{\lambda}
\newcommand{\sz}{\hat{s}_{z}}
\newcommand{\sy}{\hat{s}_y}
\newcommand{\sx}{\hat{s}_x}
\newcommand{\sio}{\hat{\sigma}_0}
\newcommand{\six}{\hat{\sigma}_x}
\newcommand{\siz}{\hat{\sigma}_{z}}
\newcommand{\siy}{\hat{\sigma}_y} 
\newcommand{\vhsig}{\vec{\hat{\sigma}}}
\newcommand{\hsig}{\hat{\sigma}}
\newcommand{\hH}{\hat{H}}
\newcommand{\hU}{\hat{U}}
\newcommand{\hA}{\hat{A}}
\newcommand{\ta}{\tilde{\alpha}}
\newcommand{\hB}{\hat{B}}
\newcommand{\hC}{\hat{C}}
\newcommand{\hD}{\hat{D}}
\newcommand{\hV}{\hat{V}}
\newcommand{\hW}{\hat{W}}
\newcommand{\hK}{\hat{K}}
\newcommand{\hX}{\hat{X}}
\newcommand{\hM}{\hat{M}}
\newcommand{\hN}{\hat{N}}
\newcommand{\te}{\theta}
\newcommand{\vze}{\vec{\zeta}}
\newcommand{\vet}{\vec{\eta}}
\newcommand{\vx}{\vec{\xi}}
\newcommand{\vc}{\vec{\chi}}
\newcommand{\hro}{\hat{\rho}}
\newcommand{\vro}{\vec{\rho}}
\newcommand{\hR}{\hat{R}}
\newcommand{\half}{\frac{1}{2}}
\renewcommand{\d}{{\rm d}}
\renewcommand{\top }{ t^{\prime } }
\newcommand{\oz}{{(0)}}
\newcommand{\sint}{{\rm si}}
\newcommand{\cint}{{\rm ci}}
\newcommand{\de}{\delta}
\newcommand{\ep}{\varepsilon}
\newcommand{\De}{\Delta}
\newcommand{\eps}{\varepsilon}
\newcommand{\si}{\hat{\sigma}}
\newcommand{\om}{\omega}
\newcommand{\tr}{{\rm tr}}
\newcommand{\e}{\hat{\eta }}
\newcommand{\ha}{\hat{a}}
\newcommand{\gam}{\gamma ^{(0)}}
\newcommand{\BEQ}{\begin{equation}}
\newcommand{\EEQ}{\end{equation}}
\newcommand{\BEA}{\begin{eqnarray}}
\newcommand{\EEA}{\end{eqnarray}}
\newcommand{\sph}{spin-$\frac{1}{2}$ }
\newcommand{\ad}{\hat{a}^{\dagger}}
\newcommand{\add}{\hat{a}}
\newcommand{\spp}{\hat{\sigma}_+}
\newcommand{\smm}{\hat{\sigma}_-}
\newcommand{\fin}[1]{|\phi^{\pm}_{#1}\rangle}
\newcommand{\finp}[1]{|\phi^{+}_{#1}\rangle}
\newcommand{\finm}[1]{|\phi^{-}_{#1}\rangle}
\newcommand{\lfin}[1]{\langle \phi^{\pm}_{#1}|}
\newcommand{\lfinp}[1]{\langle \phi^{+}_{#1}|}
\newcommand{\lfinm}[1]{\langle \phi^{-}_{#1}|}
\newcommand{\lfinn}[1]{\langle\phi^{\pm}_{#1}|}
\newcommand{\z}{\cal{Z}}
\newcommand{\RI}{\hat{{\cal{R}}}_{0}}
\newcommand{\Rt}{\hat{{\cal{R}}}_{\tau}}
\newcommand{\nn}{\nonumber}

\title{Nonlinear Dynamic Models of Conflict via Multiplexed Interaction Networks}
\author{Gerardo Aquino$^{1,2}$, Weisi Guo$^{1,3}$, Alan Wilson$^1$}
\affiliation{$^{1}$ The Alan Turing Institute}
\affiliation{$^{2}$ University of Surrey}
\affiliation{$^{3}$ University of Warwick}%

\begin{abstract}
The risk of conflict is exasperated by a multitude of internal and external factors. Current multivariate analysis paints diverse causal risk profiles that vary with time. However, these profiles evolve and a universal model to understand that evolution remains absent. Most of the current conflict analysis is data-driven and conducted at the individual country or region level, often in isolation. Consistent consideration of multi-scale interactions and their non-linear dynamics is missing. 

Here, we develop a multiplexed network model, where each city is modeled as a non-linear bi-stable system with stable states in either war or peace. The causal factor categories which exasperate the risk of conflict are each modeled as a network layer. We consider 3 layers: (1) core geospatial network of interacting cities reflecting ground level interactions, (2) cultural network of interacting countries reflecting cultural groupings, and (3) political network of interacting countries reflecting alliances. Together, they act as drivers to push cities towards or pull cities away from war. 

Using a variety of data sources that overlap consistently between $2002-2016$, we show, that our model is able to correctly predict the transitions from war to peace and peace to war with F1 score of 0.78 to 0.92 worldwide at the city scale resolution. As many conflicts during this period are auto-regressive (e.g. the War on Terror in Afghanistan and Iraq, the Narco War across the Americas), it is important to show that we can predict the emergence of new war or new peace. Indeed, we demonstrate successful predictions across a wide range of conflict genres and we go on to perform causal discovery by identifying which model component led to the correct prediction. A variety of diverse conflict case studies were generated: Somalia (2008-13), Myanmar (2013-15), Colombia (2011-14), Libya (2014-16), and Yemen (2011-13). In all these cases, we identify  the set of most likely causal factors  and how it may differ across a country and change over time.
\end{abstract}

\maketitle
\section{Introduction}
Whilst conflict remains one of the greatest economic damages and barriers to the UN sustainable development goals, a systematic understanding of its time-varying causal factors on a global scale remains absent. Yet, understanding conflict and making useful predictions is vital to informing peacekeeping and diplomatic activities \cite{Guo18, Hegre19}. Examples include prioritising peacekeeping resources, identifying causal factors in peace negotiations, and informing interventions such as embargo and political pressure \cite{Gleditsch05, owidwarandpeace, PRIO18, Spagat18}.

There are many forms of conflict and its nature has evolved in recent history. Since the end of the Cold War, the frequency and severity of conventional state conflict has decreased \cite{Findley12, Clauset18}. On the other hand, there has been a steady increase in political violence by non-state actors and by international criminal organisations. Protracted political violence \cite{Halvard02, Halvard09, Toft05} plague many regions in the world across different genres \cite{Raleigh15}, ranging from the War on Terror in the Middle East, the Narco-War in the Americas, persistent ethnolinguistic violence in South Asia, to intermittent civil wars in Sub-Saharan Africa. 

\subsection{Current Approaches}
Current conflict modeling can be largely categorised as statistical regression or mechanical model fitting. 

\subsubsection{Statistical Models}
Statistical models started in the 1940s with general laws on the frequency and severity of major wars by Lewis Fry Richardson \cite{Richardson41}, and this line of enquiry has been extended to modern conflicts \cite{Spagat18, Bohorquez09, Clauset10, Clauset13} as well as urban terrorism events \cite{Guo19}. Other temporal trend analysis have attempted to discover cycles in human decision processes (e.g. memory of wars) and climate cycles (e.g. dry-wet cycles affect agricultural output) , \cite{Klingberg66, Denton68}. Whilst these trends can inform us of the both the nature (e.g. memoryless) and likelihood of new events, they do not identify a wide range of causal factors (e.g. climate change~\cite{Hsiang11, Hsiang2013, Bai2011}, ethnolinguistic tensions \cite{Weidmann09}, and natural resources \cite{Halvard09}) and are often only accurate at very rough resolutions. Detailed multi-variate analysis can confirm or discover new insights into ongoing violence \cite{Hegre17, Hegre19}, but the profile of factors will vary across regions and evolve over time. Whilst such analysis finds use in quasi-static periods of conflict, e.g. the auto-regressive War on Terror, where a self-consistent behaviour is present; it has trouble predicting the emergence of new crisis (e.g. Ukraine conflict) or the sudden end of a protracted conflict (e.g. Colombia peace agreement). Indeed, even current machine learning approaches, which are effectively automated regression models, perform well for white swan conflict events and poorly for black swan events \cite{Guo18}. The open challenge for statistical models remain the non-linear time-varying nature of conflict.

\subsubsection{Mechanistic Models}
Mechanistic models are useful in describing the non-linear dynamics of conflict. Classical examples include applying the Hawkes process, which is a self-excitation behavioural modeled \cite{Fry16}. Here, the assumption is that successful attacks encourage both the attacker to repeat their success, but also exacts counter-measures by the security forces. This has been successfully implemented for IED attacks in Northern Ireland, Iraq, and Afghanistan. Expanding this model to a spatial diffusion process, recent work on the Afghan War Diaries have integrated point conflict processes with a stochastic integral kernel \cite{Afghan12, Gao13} to create a dynamic diffusion map. Other models also examine spatial correlations between conflict dynamics and infer a causation network for analysis \cite{Larralde15}. Long term agent based models based on a series of dynamics have also been applied to historical contexts \cite{Turchin_13}.

\subsection{Contribution \& Coverage}

\subsubsection{Current Limitations \& Our Novelty}
Current models are lacking in explaining several aspects of conflicts:
\begin{enumerate}
    \item how does the set of causal factors which exasperate conflict vary across space and evolve over time?
    \item why do certain locations suffer a disproportionate level of violence \cite{Krieger11} (relative to rest of the conflict region and their population size)?
    \item how can we build a quantitative model that integrates diverse causal factors ranging from geography to politics to ethnolinguistic tensions.
\end{enumerate}

Here, we attempt to resolve the above 3 shortcomings in research by the following approach:
\begin{enumerate}
    \item model each city as a bi-stable system, where it will always gravitate towards war or peace, with an unstable intermediary skirmish or risk of war state. This enables us to model the nonlinear dynamics with the simplest model (used in prevalence in ecology and biology). Causal factors then contribute to the model by pushing it towards or pulling it away from conflict.
    \item connect each city via a multi-layered spatial interaction network, whereby the network structure itself reveals vulnerable contentious points that we argue to be at a higher risk of violence \cite{Guo17}.
    \item integrate the above into a global framework, whereby the tuning parameters are trained on previous conflict data to: (a) make predictions, and (b) discover which set of evolving causal factors dominate.
\end{enumerate}

\subsubsection{Data Sources and Study Period}
In this paper, we consider both non-state terrorism and conventional warfare from 2002 to 2016, for which there is spatially accurate reporting of violent events based on the Global Terrorism Database (GDT). 
We attempt to model 3 general measures of causality for conflict: (1) current and recent conflict history, (2) cultural affinity, and (3) political relation. We know that many conflict in sub-Saharan Africa are extremely regressive and therefore point (1) is important. The data input for (1) uses the previous state of conflict and the neighbouring current states of conflict. In (2), we account for cultural diversity  at the country level by using a database of the religious background of the population of each country \cite{Religion}. In (3), we use a variety of data sources to infer the nature of the country's political relation with other countries. We first use data from \cite{SIPRI} to check if there are formal alliance relationships (e.g. NATO, EU) and then go on to check if there are arms trade in recent years. Whilst positive relations are reflected in this manner, inferring negative relations is difficult. As such, we infer that the absence of any political treaties or military trade implies a frosty or hostile relationship. 

Resolution wise, we work per year and for (1), the resolution is at a city and surround region level, and for (2) and (3) it is at the country level, which is then uniformly associated with each city in those countries.

\subsubsection{Organisation of Paper}
The paper is organised as following. In section II we introduce the network, how it is derived and its multiplex structure. In section III we introduce the  dynamical model we implement on the network. In section IV and V we show results of simulations and predictions  based on data and discuss their significance in relation to other methods. In section VI we consider the role of network structure in identifying high conflict areas. In section VII we consider specific cases and show how the model allows one to trace causal links to the emergence of conflict events, followed by final remarks.

\section{Multiplexed Interaction Network}
\begin{figure}[h!]
\includegraphics[height=8.7 cm,width=9.9 cm, angle=0]{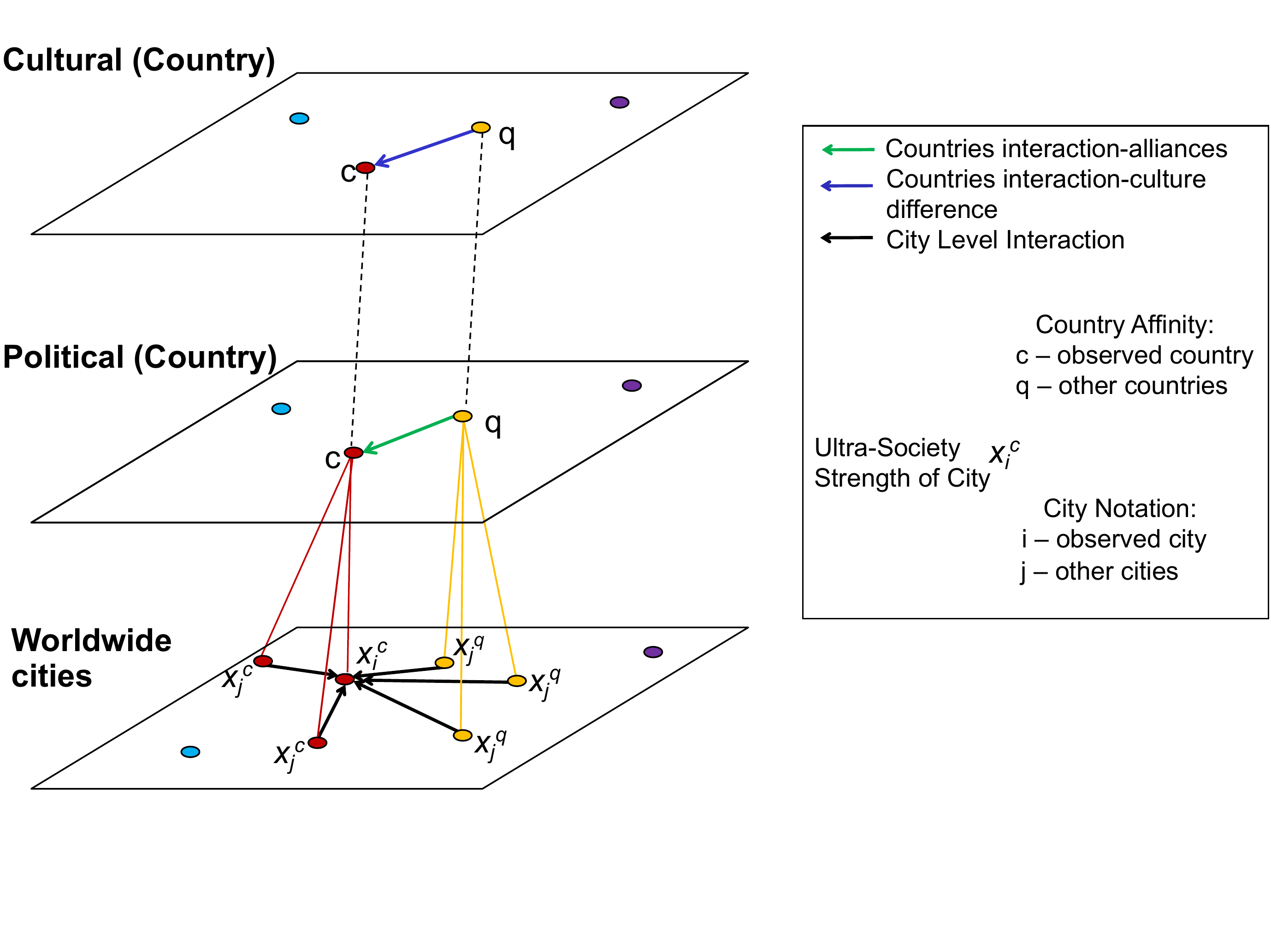}

\caption{
Three-layers structure of  the multiplex-network  on which the model is implemented.
}
\label{fig1}
\end{figure}

We consider a set of $N=7323$  cities worldwide distributed.
We implement on this set a multiplexed network consisting of 3-Layers, as illustrated Figure \ref{fig1}:  from bottom up we have  core Geography,   strength of political relationships and a cultural layer accounting for the cultural differences between populations of different countries. 

The nodes in the geographic layer represent cities/towns and the links in the layer represent interactions between nodes.  Each node is also attached to its country and countries are linked via political  and cultural interactions. Within a layer, the links between different nodes represent  interactions. The state of a node, measured by the variable $x^c_i$,  with $i$  denoting the city and  $c$ its country affiliation, is the focus of the model. This variable represents a well-being factor, which in ecology is commonly population, but in our case, it  is the ultrasociety power, i.e., the strength of the cooperation, which, as originally introduced in  \cite{Turchin_13}, is a key indicator of the strength of a society. Our dynamic model, introduced in Section III, follows the evolution of this variable in each node, characterised by transitions between poor (war) and optimal (peace) well-being due to interactions with the other nodes via the different layers of the multiplex structure of the network. 

\subsection{Geographic layer: Adjacency matrix}
 We follow the same procedure detailed in \cite{Guo17} and consider an entropy maximizing coupling dynamics between cities \cite{Wilson70, Wilson67} to define the adjacency matrix for the network of cities.   The entropy maximisation allows  one to derive the flows  between nodes in the network by maximising the entropy with  the natural constraints on the origin-destination matrix and an overall cost \cite{Wilson70}.
 More precisely, in a constrained system, where each city  has a finite capacity to interact with neighbours, the flow between nodes can be measured via the Boltzmann-Lotka-Volterra (BLV) model (see \cite{Wilson08} for a review). Let $F_{ij}$  be the flow between nodes $i$ and $j$ in the geographic layer and   $F$  the total possible flow on the network, the likelihood $W(F_{ij})$  of a given flow between two nodes is  \cite{Wilson70}:
 \BEQ
 \label{likd}
 W(F_{ij})=\frac{F!}{\Pi_{ij} F_{ij}}
 \EEQ
 The most likely value for $F_{ij}$ is obtained by maximising the logarithm of (\ref{likd}), which can be shown to be equivalent to maximising the entropy $\sum_{ij} F_{ij}\log F_{ij}$.
 As shown in \cite{Wilson70} maximising entropy with constraints on a global flow energy, on the total flow $X_i$ coming from a single node and
 on the total capacity of the system (depending on node capacity $Z_i$)
 leads to
 \BEQ
 F_{ij}=X_i^a \frac{Z_j^b/f(d_{ij})}{\sum_k Z_k^b/f(d_{ik})},
 \EEQ
 where $f$ is a function of the distance $d_{ij}$ between nodes.
 Setting $Z_i$ and $X_i$ equal to population $p_i$ in node $i$   reproduces the gravity law (a part from normalisation term),  widely observed  in  real space networks constrained by geographic distances (see \cite{And, Berg, GuoAir} and references therein).
 We adopt this procedure and derive the flows on the network, the weights of the adjacency matrix are then defined as the inverse of the flows.
 
 A clustering radius of 500 Km  is introduced, representing 0.5 \% of the global surface and only cities within this radius are connected. As shown in \cite{Guo17} this radius lies at the transition to percolation over the global network.
 Also, cities separated by sea of a distance larger than 50 Km, are disconnected. We limit ourselves for this paper only to land travel and short distance water travel between cities to build our network.
 Distances are evaluated using latitude and longitude data and geodesic distance. See \cite{Guo17} for further details.

\subsection{Political layer and Alliance Matrix}
We  extract political alliance relationships from one data set including  alliances between 183 countries from 1950 to 2015,  provided by the Stockholm International Peace Research Institute (SIPRI)  \cite{SIPRI} for years up to 2012 and integrated with arm trade data and data from the   Alliance Treaty Obligations and Provisions (ATOP) project  \cite{ATOP}. We assume the absence of any political treaty or arm trade as indication of hostile relationship. We use this information to build an alliance matrix $\mathcal{P}$  with  $\pm 1$ values for enemy (E) or ally (F) relationship between nodes. The political relationship in combination with the  state of the nodes   determines  the sign of the  interaction  between of two nodes in our model (see Section III and  Eqs. (\ref{runz2}), (\ref{run})), i.e. whether the interaction strengthens or destabilises the well-being of the node.


\subsection{Cultural Layer}
It is reasonable to expect   that cultural difference/similarity between countries will affect the way they interact,  respectively enhancing or  mitigating friction between countries arising from political-economical  competition.
We implement this in our  network and dynamical model  by gathering information from a dataset including statistics of religious affiliation of population for 167 countries. We build a vector for each country
with each component  of the vector representing the percentage of population for each religious affiliation. We then define a cultural distance $d_c$ between countries
as the Euclidean distance between these vectors. For the few countries where no data is available we assume  as  demographics, an average over all other countries.
We weigh with this distance the interaction term between nodes so as to increase or decrease the drag of the interaction towards conflict according to the degree of cultural difference between the nodes involved. The reasoning  behind this choice is that  the push on a node  towards war should be dampened when neighbour node belongs to countries with similar cultural background and heightened  by different cultural background. Accordingly, the push towards peace should be stronger when the cultural difference  between two interacting nodes is low and weaker when it is high.
We therefore  define  the  matrix $\mathcal{C}_{ij}$  which weighs the interaction between nodes based on the cultural difference between the populations as:
\BEQ
\label{cult}
\mathcal{C}_{ij}= \begin{cases}
     d_c(i,j), & \text{if}\ i,j \text{ enemies} \\
      1-d_c(i,j), &   \text{if}\ i,j \text{ allies}
    \end{cases}
  \EEQ
   this term will implement the effect of the cultural layer of the network on the dynamics  (see Eq. \ref{run}) exactly in the way described above.

\subsection{Datesets}
We adopt the following data sets:

World cities data:
7323 cities with their latitude, longitude, and population data from National Geospatial Intelligence Agency \cite{NSA}.  The data represents a quarter of the world's total population and includes over 2800 cities with a population over 100,000 units, yielding a sufficiently high city resolution. Each city is also tagged with its country and province affiliation.

Conflict data:  For terrorism and insurgency violence: the Correlates of War  \cite{COW} database is used, with over 180,000 conflict incidents between $2002$ and $2016$. The GTD contains the number of attacks and death-toll, ranging from small-scale attacks (1 death) to large-scale massacres (1000s dead). Most of the death-toll data is time stamped and geo-tagged (longitude and latitude).

Political alliance date:  we use arm trade and political alliance data as provided by  the Correlates of War (CoW) project \cite{COW}
integrated to fill for years from 2012 to 2016 with data from  Stockholm International Peace Research Institute (SIPRI) \cite{SIPRI} and the Rice University  Alliance Treaty Obligations and Provisions (ATOP) project  \cite{ATOP}.

Religious affiliation data:  statistics of world countries populations  according to the PEW Research Center \cite{Religion} as a measure of the cultural difference between cities of different countries.

\section{Model}
Like many ecological systems, the dynamics of population or growth power (well-being) can be  modelled by a 3rd order polynomial, which is equivalent to having a double-well potential. Therefore in each node $i$ we have
\BEQ
\label{runz}
 \frac{d x_i}{d\tau} = \left(1-\frac{x_i}{a_i}\right) \left(1+\frac{x_i}{c_i}\right)\left(x_i-b_i\right)  + \eps H_i
 \EEQ
where we drop the country index, as compared to notation in Fig. \ref{fig1} in order to lighten notation. In the following lower indexes $i,j,k$ refer to nodes in the geographic layer of the network.
We use $\tau$ to indicate the dynamical time on which  Eqs. (\ref{runz}) 
are run until all nodes reach  a steady-state,  to differentiate it from the real time to which data refer.
The rate of  well-being/growth power in a node is determined by: (i) a regressive component (stronger nodes tend to get stronger), 
(ii) a capacity limit ( a node cannot be stronger than natural resources dictate), 
(iii) a critical mass threshold (must be of a minimal size to grow), and 
(iv) a constant 'field' term (varying from year to year) related to generic external factors. In this particular third order dynamics, there are up to three equilibrium points, two stable  and one unstable in between. 
The "external field" term may  take in account other economic parameters, trade or resource availability which can shift the equilibrium of the nodes. Interpreting this, we may understand the following:

-When there is only one positive equilibrium point, the society is peaceful and strong. Possibly leading to it being a pure aggressor.

-When there are three equilibrium points, the society can enter a state of uncertainty (middle equilibrium) and then war (undesirable equilibrium).

-If the undesirable equilibrium is positive, peace is possible. If it is negative, it will devastate the population to the point of no recovery.
 To induce transition from one stable equilibrium to another, we use external forces from other cities, whereby the sign of the forcing ($\pm $) is determined by national relationships and current node states.
 \begin{figure}[h!]
\includegraphics[height=8. cm,width=15.8 cm, angle=0]{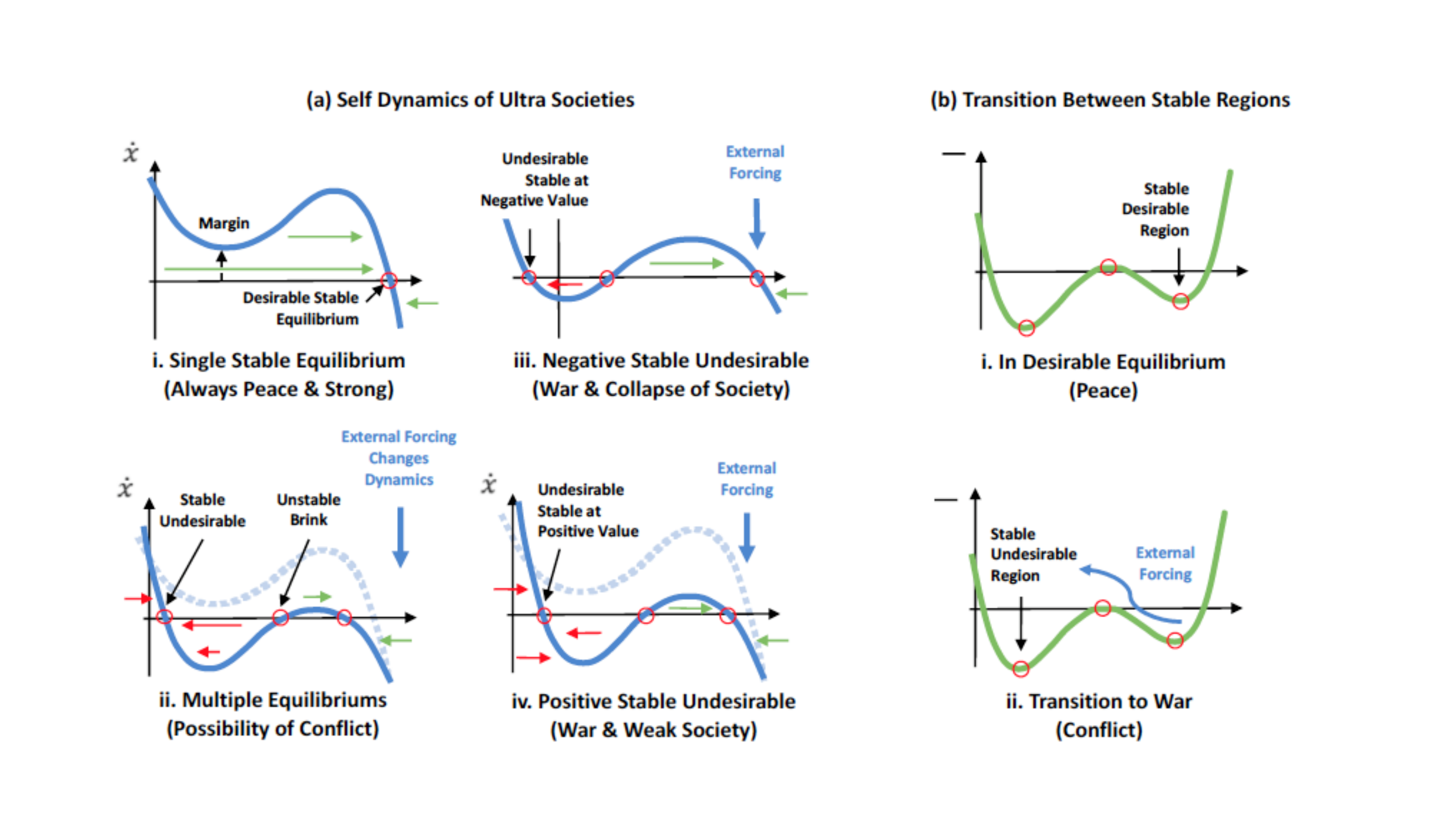}
\label{fig2}
\caption{Driving dynamics in Eq. 1 (a) and relative potential with induced transitions (b).}
\end{figure}

In  the self-dynamic equation (\ref{runz}), $a_i$ controls the size of the largest equilibrium, $b_i$ ($b_i<a_i$) controls the unstable brink equilibrium and $c_i$ the smallest equilibrium.  For systems with a favourable $H_i$ value, there exists only one positive equilibrium that is above $a_i$, meaning society will flourish beyond its natural capacity. For systems with a negative $H_i$ value, there may exist only one negative equilibrium value, meaning that society will vanish. 
 
 \subsection{Interaction}
 The coupling dynamics that  reinforces or weakens the well-being of a city due to interaction with neighbours  is given by: 
  \BEQ
  \label{runz2}
 \frac{d x_i}{d\tau} = -\gamma_i  \frac{\sum_j  A_{ij} x_i S_{ij}\mathcal{P}_{ij}f(d_{ij}) x_j}{\sum_k x_i A_{ik}x_k f(d_{ik})}
\EEQ
with $A_{ij}$ the adjacency matrix,  $\mathcal{P}_{ij}$ the matrix of political alliances, with value $\pm 1$ indicating  friend(F)/enemy (E) and
$f(d_{ij})$ a function of the geographical distance. The operator $S_{ij}$
is introduced to ensure that the sign of interaction reflects the  political alliance and the state of each interacting couple of nodes $i,k$. Namely, if node $k$ is in a war state, here defined as $x_k<0$,  it will tend to drag an ally $i$ to war as well, that is, it will push node $i$ towards negative values $x_i<0$. A neighbour enemy node will have the same effect. In all other cases, $S_{ik}$ ensures that the effect of node $k$ on $i$ is to push the node $i$ towards a peace state ($x_i>0$).  These conditions on $S_{ik}$ are  easily implemented algorithmically in our simulations.
For the distance function $f(d)$ we adopt an inverse square law. $\gamma_i$ is a positive parameter vector which accounts for local variability in the strength of the interaction.
\subsection{Combined Dynamics}
Combining self- and coupling-dynamics yields from Eqs.(\ref{runz}) and (\ref{runz2}): 
  \BEQ
    \label{run}
  \frac{d x_i}{d\tau} =  \left(1-\frac{x_i}{a_i}\right) \left(1+\frac{x_i}{c_i}\right)\left(x_i-b_i\right) +\eps H_i -\gamma_i  \frac{\sum_j  x_i A_{ij} \mathcal{P}_{ij} S_{ij}\mathcal{C}_{ij} f(d_{ij}) x_j}{\sum_k x_i  A_{ik}x_k f(d_{ik})}
\EEQ
where we have  added the cultural layer  $\mathcal{C}_{ij}$  as defined in (\ref{cult}), weighing cultural differences between nodes.
Depending on the sign of $\mathcal{P}_{ij}$  and nodes states the coupling dynamic can either lower the risk margin or increase the risk margin of war. 
We track the number of equilibria for each city and whether it is in a high or low equilibrium state during a long simulation time (to achieve steady state). Low equilibrium  corresponds to conflict.

\section{Model results on  Aggregate Data}
\begin{figure}[h!]
\begin{subfigure}{7.8cm}
\hspace{-7.cm}  \includegraphics[width=14.5cm]{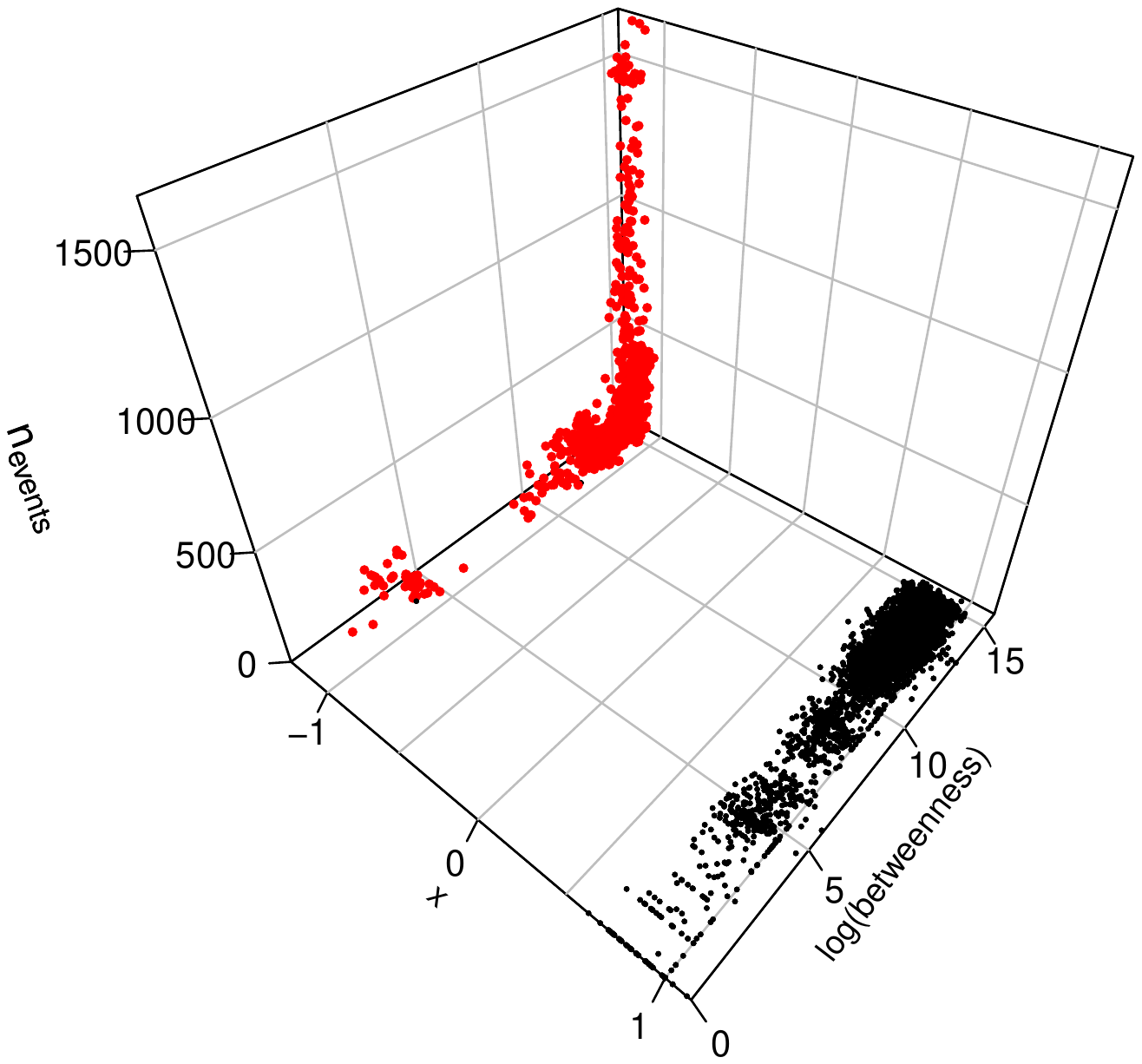}
  \end{subfigure}
  \begin{subfigure}{5.2 cm}
    \includegraphics[width=6.8cm]{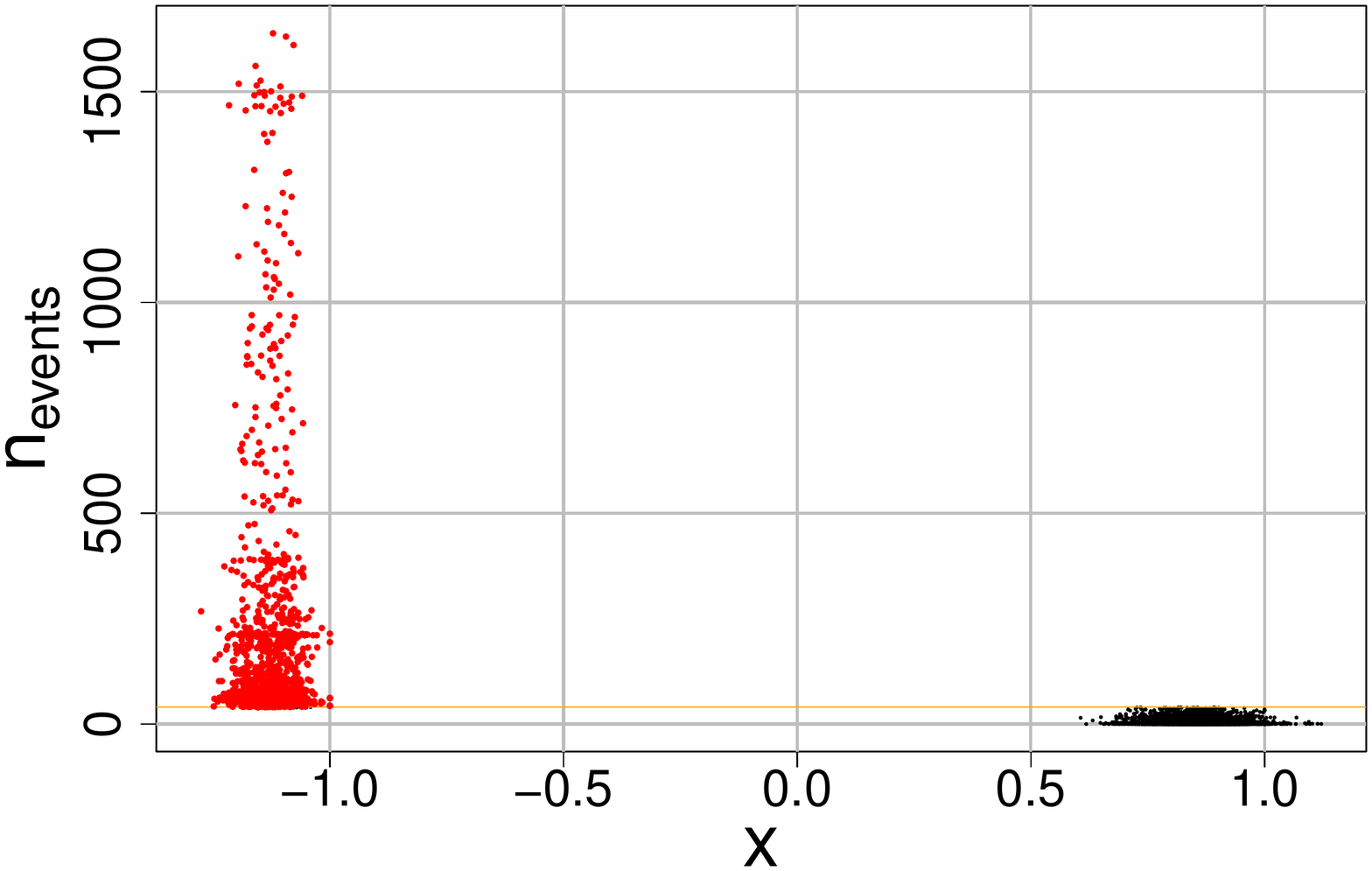}
    \includegraphics[width=6.8cm]{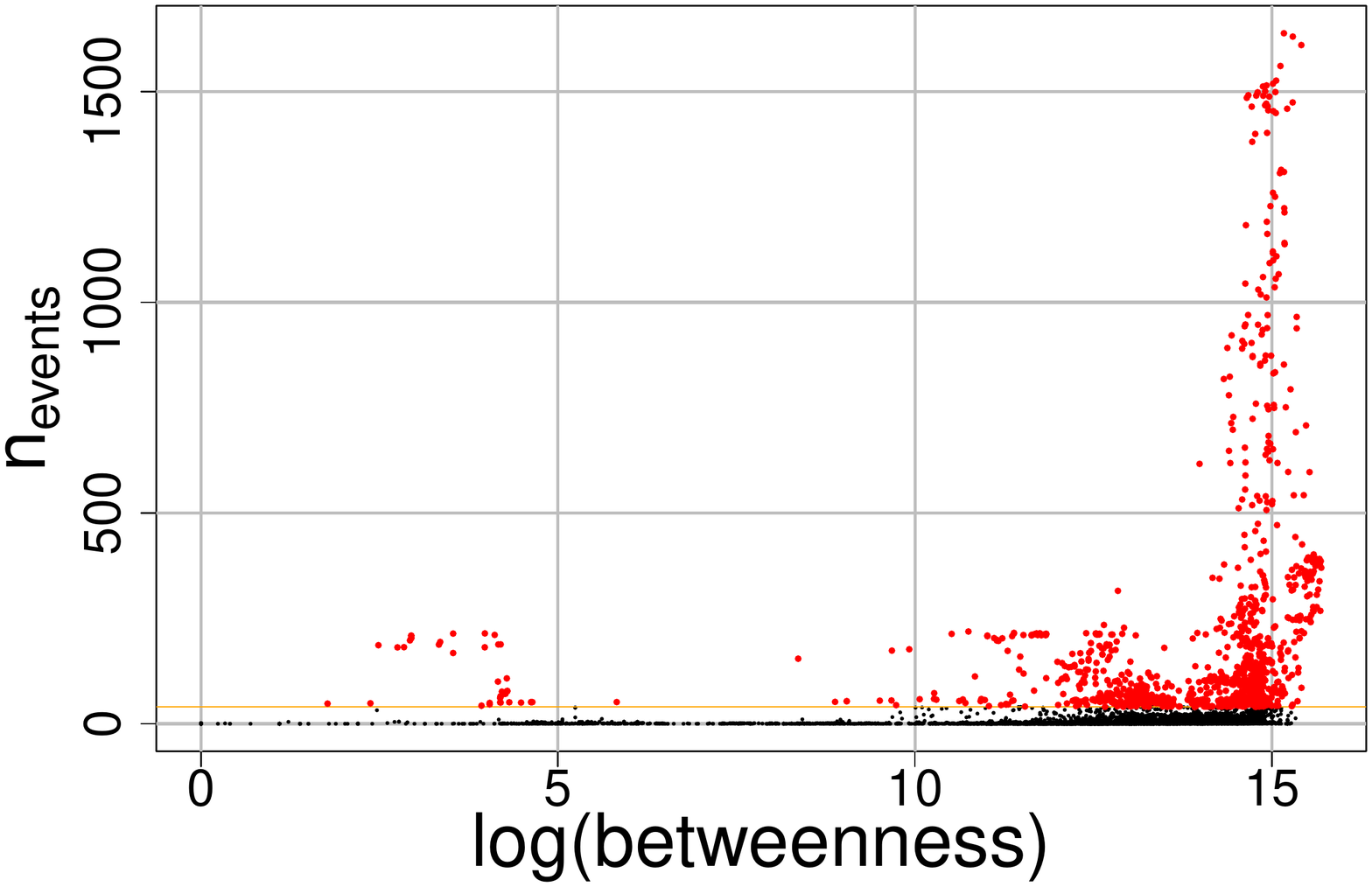}
  \end{subfigure}
\caption{(Left)Simulation of model based on aggregate data $2002-2016$. Node coordinates , betweenness and number of conflict events are reported respectively on $x$,$y$ and $z$ axis.  Maximum F1 score, calculated with respect to the aggregate conflict data of same period, is 0.99 reached for threshold value of 40 number of events per year.Red dots represent areas with number of attacks per year  $>$ threshold  value. Black dots areas with lower number of  events. (Right)2D projection on x-z plane (above) and z-y plane (below) of same plot. Orange line marks threshold on number of events}
\label{fig3}
\end{figure}

We run simulations of  Eq.  \ref{run} for our set of 7323 cities, implemented via   Euler standard integration method and allow the dynamics to reach steady state.
We consider here the case of $\eps=0$  and leave for future development the possibility of using other date for the impact of economic or other resource factors to the nodes dynamics. We chose as well parameters  $a_i=1$ and $b_i=0$  and $c_i=1$for all nodes. With this choice the potential profile in Eq. (\ref{runz}) becomes a double-well  with bottoms at $x=-1$ and $x=1$ respectively.
Figure \ref{fig3}  shows the simulations at steady state in a 3D space on aggregate data from 2002 to 2016, where the $x$ axis is the well-being value according to the dynamics (\ref{run}). On the $y$ axis we plot in logarithmic scale the value of the betweenness at each node and on the $z$ axis the number of conflict events per year in each node (averaged over the period 2002-2016).
We aggregate in each node the number of events occurring per year in an aera of radius $R=500 Km$ around the node.  Our model classifies on negative $x$ 
conflict areas and peaceful areas on positive $x$. 
The plot visually shows in red conflict areas with a number of conflict events per year larger than a threshold which is determined by maximising the F1 score. This value corresponds to $40$ events
which coincides with the average value of number events per year and per node in our data set. We adopt in the following this threshold as discriminant of war/peace state in a node.
 From  Figure \ref{fig3} one can already see how high conflict areas, those with higher number of events per year, tend to accumulate on  large value of $y$, i..e. in the high betweenness areas. We will see that adding  betweenness as a classifier increases the model predictive power of high conflict areas.

\section{Model predictions after training}

We then  test the predictive power of our model. We train the model  on the previous available years prior to the year on which we want to make predictions.  We therefore estimate the coefficients $\gamma_i$ 
based on the training examples given by the data  and run the model with these learnt parameters to make prediction for the chosen year. In Fig. 4  and Fig. 5 we plot  the results for year 2016 and 2015 respectively, based on  training on  previous years starting from 2002 and considering as threshold for a war/peace state the value of 40 events per year. Initial conditions for the dynamical equations are taken
based on data of previous year and mapped, based on the threshold of 40 events, on the coordinate space of the dynamical variables $x_i$ for the nodes of the network.  
If the average value of events  on the training years in a given node is smaller/larger than the threshold and the number of events in the year considered in the given node is larger/smaller than the threshold, we mark the occurrence of a new war/peace event in that node.  In the plot red and green dots represent respectively such new conflict and peace areas arising in 2016 as compared to training. While blue and black dots represent conflict and peace areas which did not change state compared to training.
Table \ref{ttab1}  reports the  performance of the model based on different measures  in the case of binary classification war/peace of each node state.  These measures are defined as:

\BEA
\nn \text{Accuracy}&=&\frac{ \text{number of correct predictions}}{\text{total number of predictions}}\\
\nn\text{Recall}&=&\frac{ \text{number of  true positives}}{\text{number of false positives}+\text{number of true positives}}\\
\nn\text{Precision}&=&\frac{ \text{number of  true  positives}}{\text{number of false negatives}+\text{number of true positives}}\\
\nn\text{F1}&=&2\frac{ \text{Precision} \times \text{Recall}}{\text{Precision}+\text{Recall}}
\EEA

\begin{figure}[h!]

\begin{subfigure}{7.8cm}
\hspace{-7.cm}  \includegraphics[width=14cm]{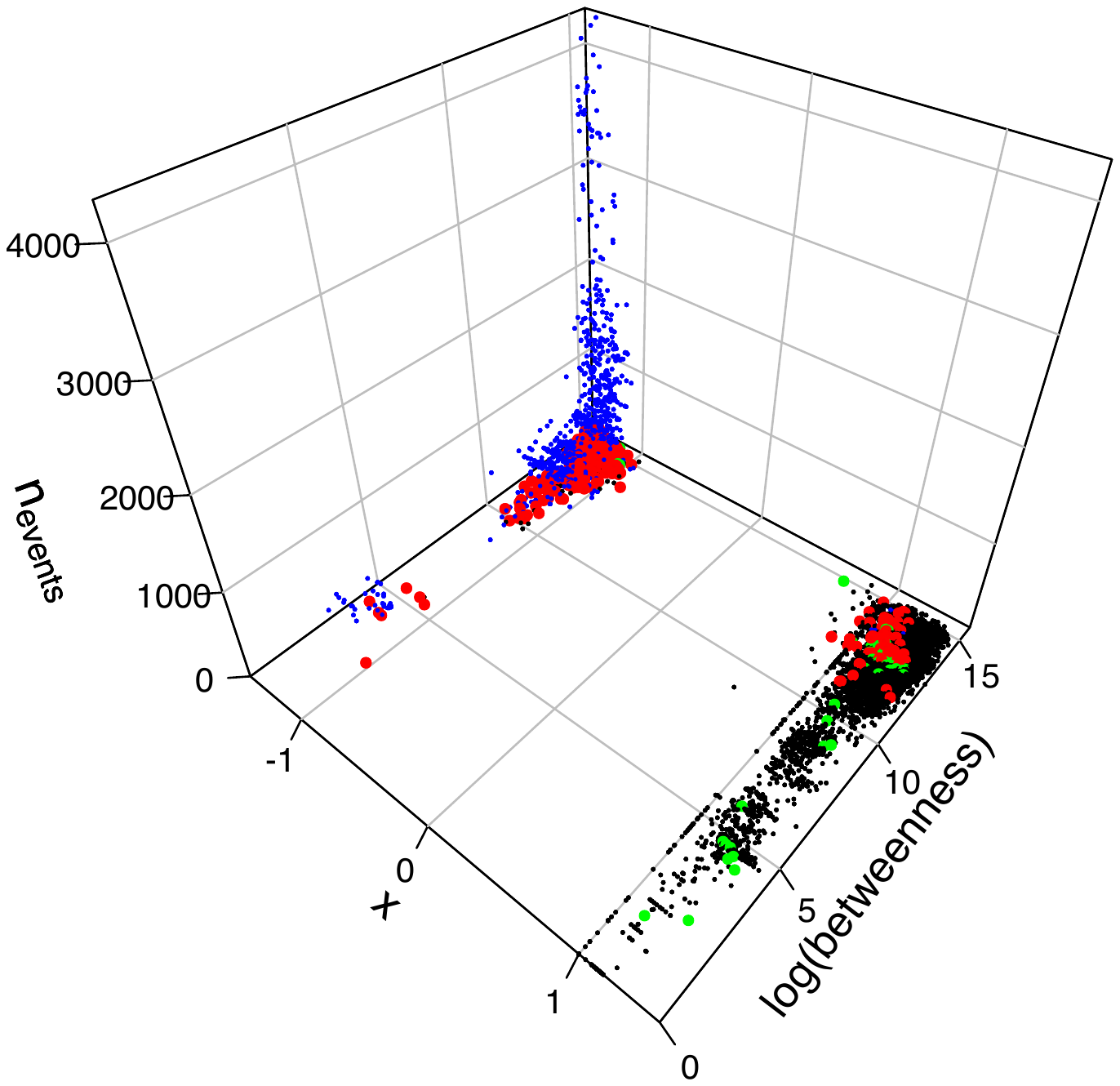}
  \end{subfigure}
  \begin{subfigure}{5.2 cm}
    \includegraphics[width=6.8cm]{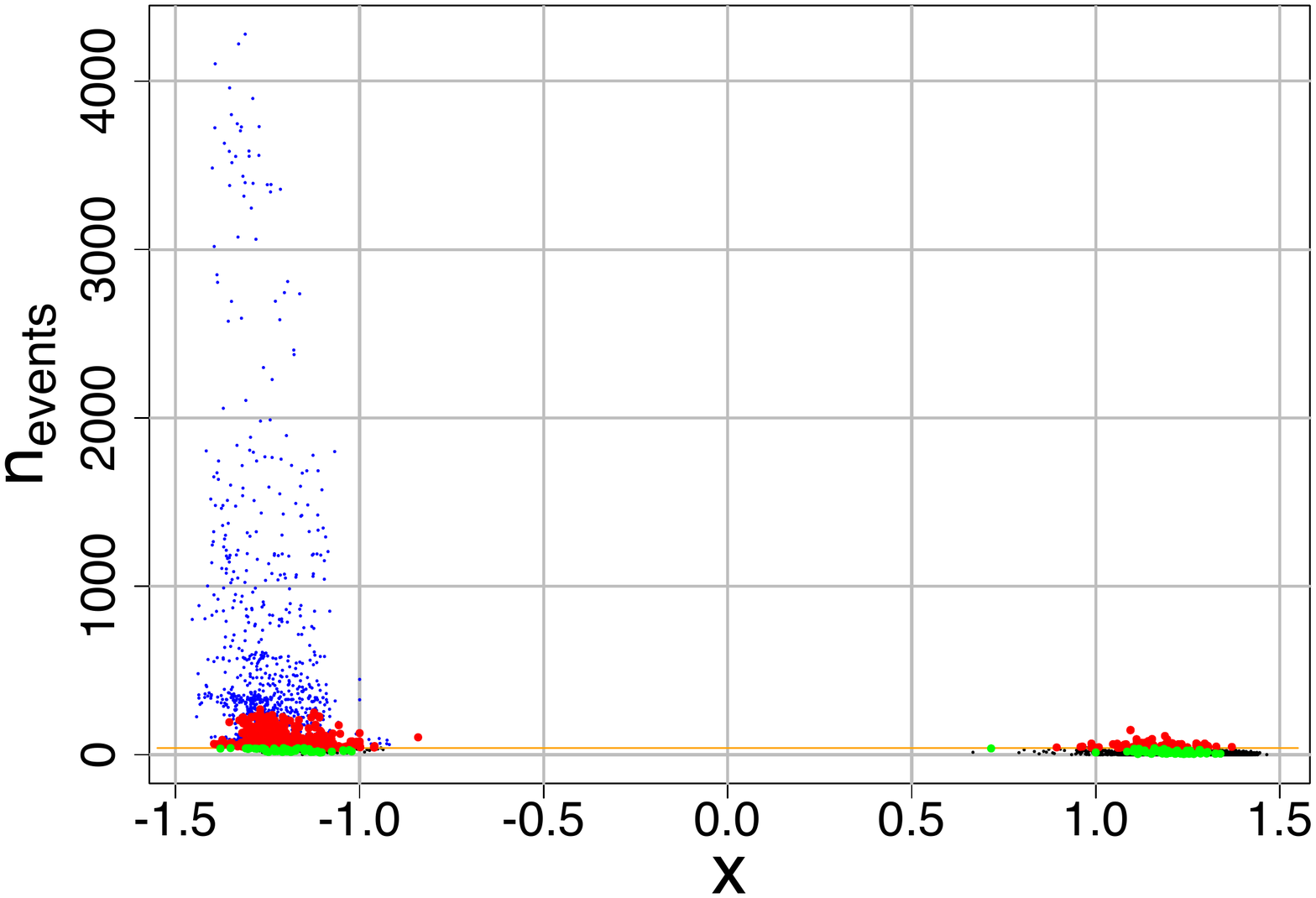}
     \includegraphics[width=6.8cm]{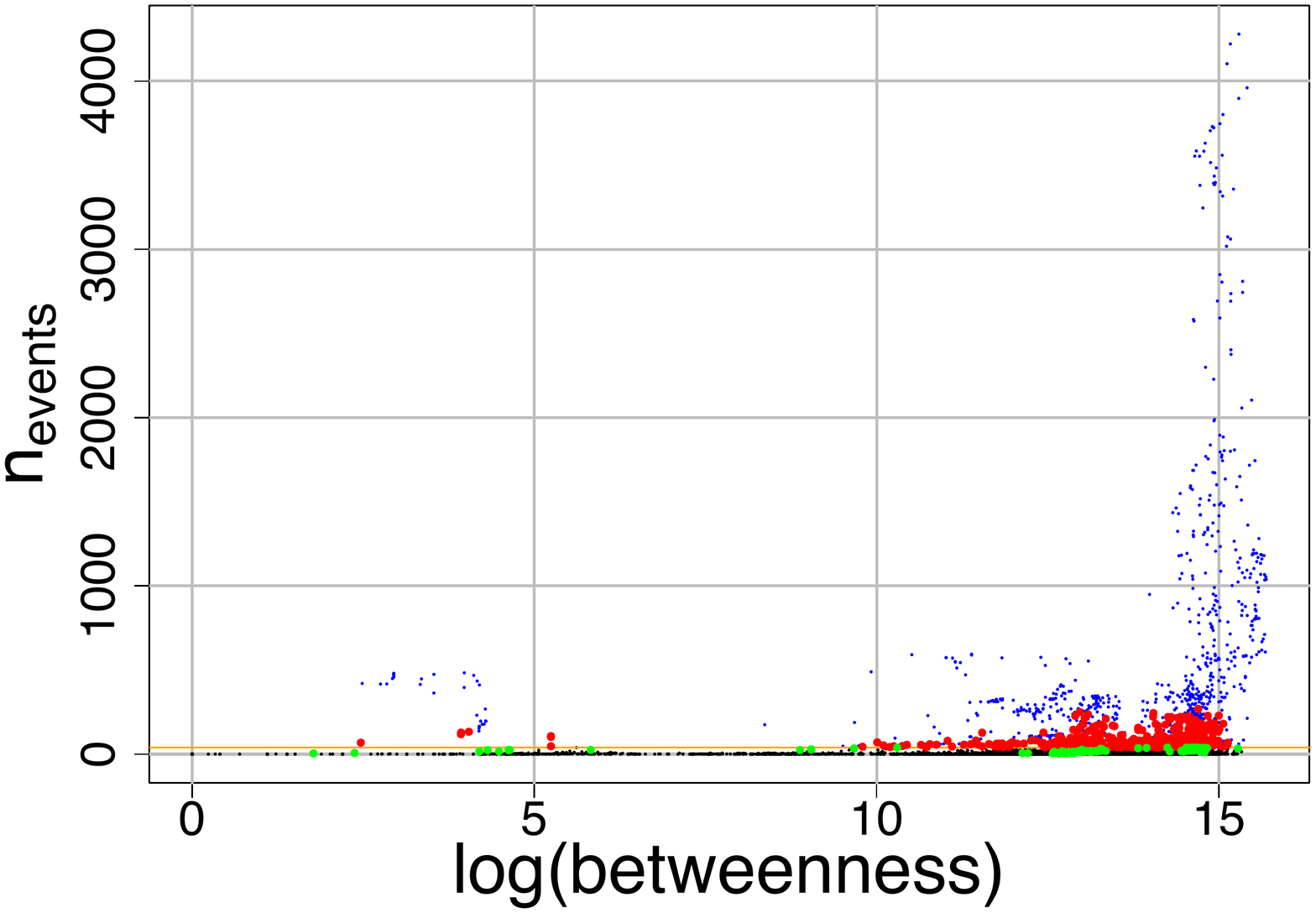}
  \end{subfigure}
\caption{Simulation for year $2016$ run after training over years $2002-2015$. (Left) Node coordinates , betweenness and number of conflict events are reported respectively on $x$,$y$ and $z$ axis. Bigger red and green dots indicate areas which have shifted respectively to conflict or peace state as compared to previous year.  Small blue and black dots indicate respectively ares of conflict and peace, which have not changed  state as compared to previous year (see Table  \ref{ttab1} , \ref{tab2} for a quantitative analysis). (Right) 2D projection on x-z plane (above) and z-y plane (below) of same plot. Orange line marks threshold on number of events.} 
\label{fig4}
\end{figure}

\begin{table}[h!]
\begin{tabular}{rlllllllllllll}

  \hline
 & measure/year & 2005 & 2006 & 2007 & 2008 & 2009 & 2010 & 2011 & 2012 & 2013 & 2014 & 2015 & 2016 \\ 
  \hline
1 & Accuracy & 0.98 & 0.98 & 0.98 & 0.98 & 0.95 & 0.96 & 0.95 & 0.95 & 0.96 & 0.93 & 0.92 & 0.96 \\ 
  2 & F1 & 0.81 & 0.92 & 0.89 & 0.78 & 0.85 & 0.78 & 0.78 & 0.85 & 0.86 & 0.81 & 0.82 & 0.91 \\ 
  3 & Recall & 0.99 & 0.96 & 0.96 & 0.98 & 0.78 & 0.89 & 0.72 & 0.99 & 0.93 & 0.9 & 0.88 & 0.88 \\ 
  4 & Precision & 0.68 & 0.89 & 0.83 & 0.64 & 0.94 & 0.7 & 0.85 & 0.75 & 0.8 & 0.73 & 0.77 & 0.95 \\ 
   \hline
\end{tabular}
\caption{Accuracy, F1 score, recall and precision for predictions based on model after training over previous years. Threshold for war/peace discrimination is 40 events}
\label{ttab1}
\end{table}

The results in Table \ref{ttab1}  show a good performance of our model in overall predicting the states of the nodes of the network. More crucial is to predict the new
events. In Table II we show how the  model performs on these events, showing an average total accuracy over the  years  of more than $55\%$.

\begin{table}[h!]
\begin{tabular}{ |p{3cm}||p{1cm}|p{1cm}|p{1cm}|p{1cm}|p{1cm}| p{1.cm}|p{1cm}|p{1cm}|p{1cm}|p{1cm}|p{1cm}| p{1cm}| }
 \hline
 \multicolumn{13}{|c|}{Model performance on new conflict/peace occurrences} \\
 \hline
  year & 2005 & 2006 & 2007 & 2008 & 2009 & 2010 & 2011 & 2012 & 2013 & 2014 & 2015 & 2016 \\ \hline
green dots in $x>0$ &  73&49&54&1&1&42&50&63&53&52&57& 57 \\
green dots in $x<0$  &  0&7&5&2&32&59&67&0&7&34&16&48\\
 \% correct peace  &100\%&  87\%&  91\%& 33\% &3\% &41\% & 43\% &100\% &88\% & 60\% &78\% &54\%\\
red dots in $x<0$  &66 & 66&  62 & 102&  155 & 123&155&187&255&278&362&479\\
red dots in $x>0$   &100&49&  89 &271 & 45  & 273 & 75 & 218 & 193 & 369 &337&65\\
 \% correct war  &40\%& 57 \%&  41\%& 27\% &78\%& 31\% &67\% &47\% & 57\% &43\% &53\%&89\%\\
\% correct total  &58\%& 67 \%&  55\%& 27\% &67\%& 33\% &59\% &53\% & 61\% &45\% &55\%&83\%\\

 \hline
\end{tabular}
\caption{Performance of the model on predicting occurrence of new changes of node state to conflict or peace for years 2005-2016.}
\label{tab2}
\end{table}

\begin{figure}[h!]

\begin{subfigure}{7.8cm}
\hspace{-7.cm}  \includegraphics[width=14cm]{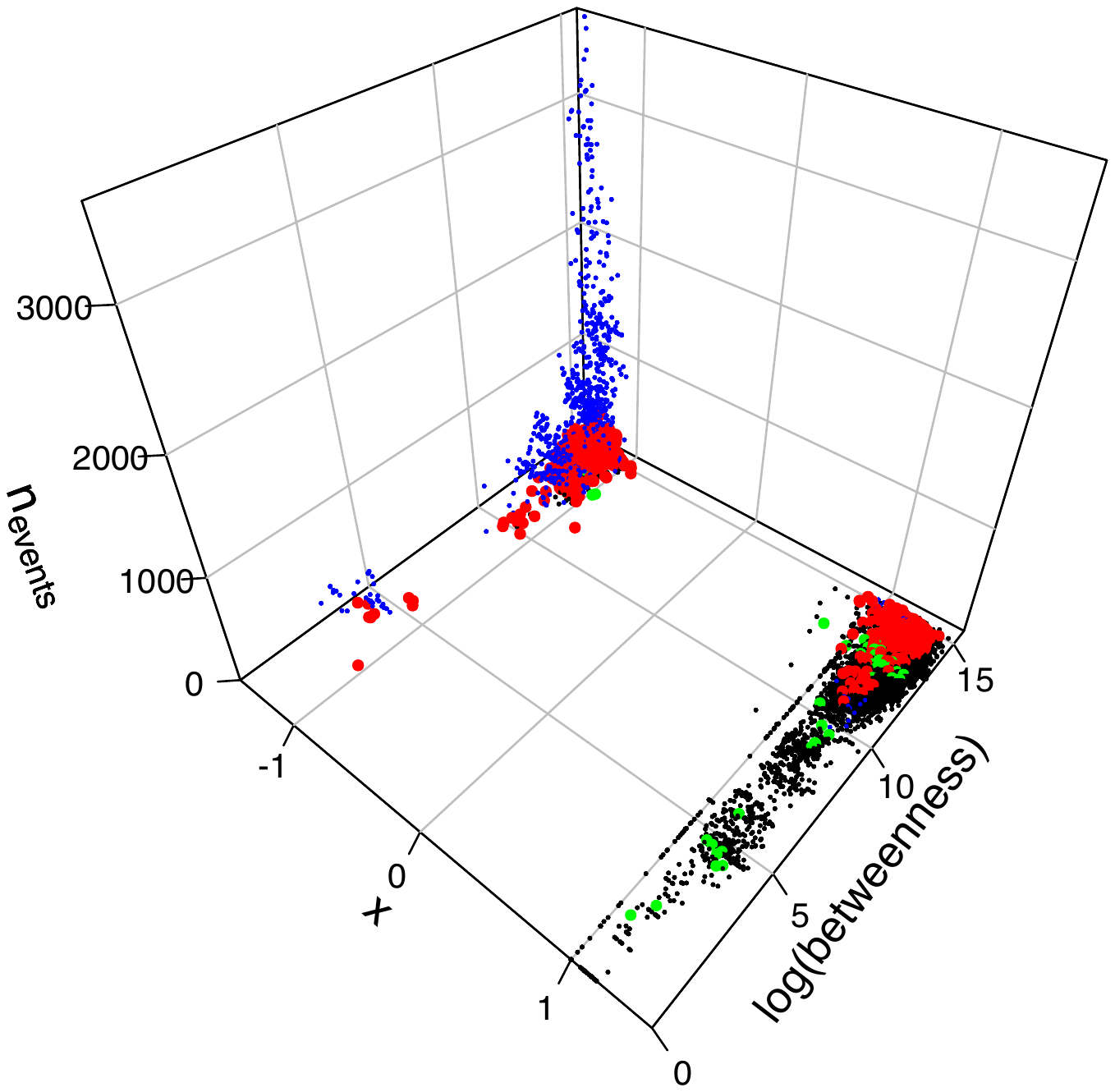}
  \end{subfigure}
  \begin{subfigure}{5.2 cm}
    \includegraphics[width=6.8cm]{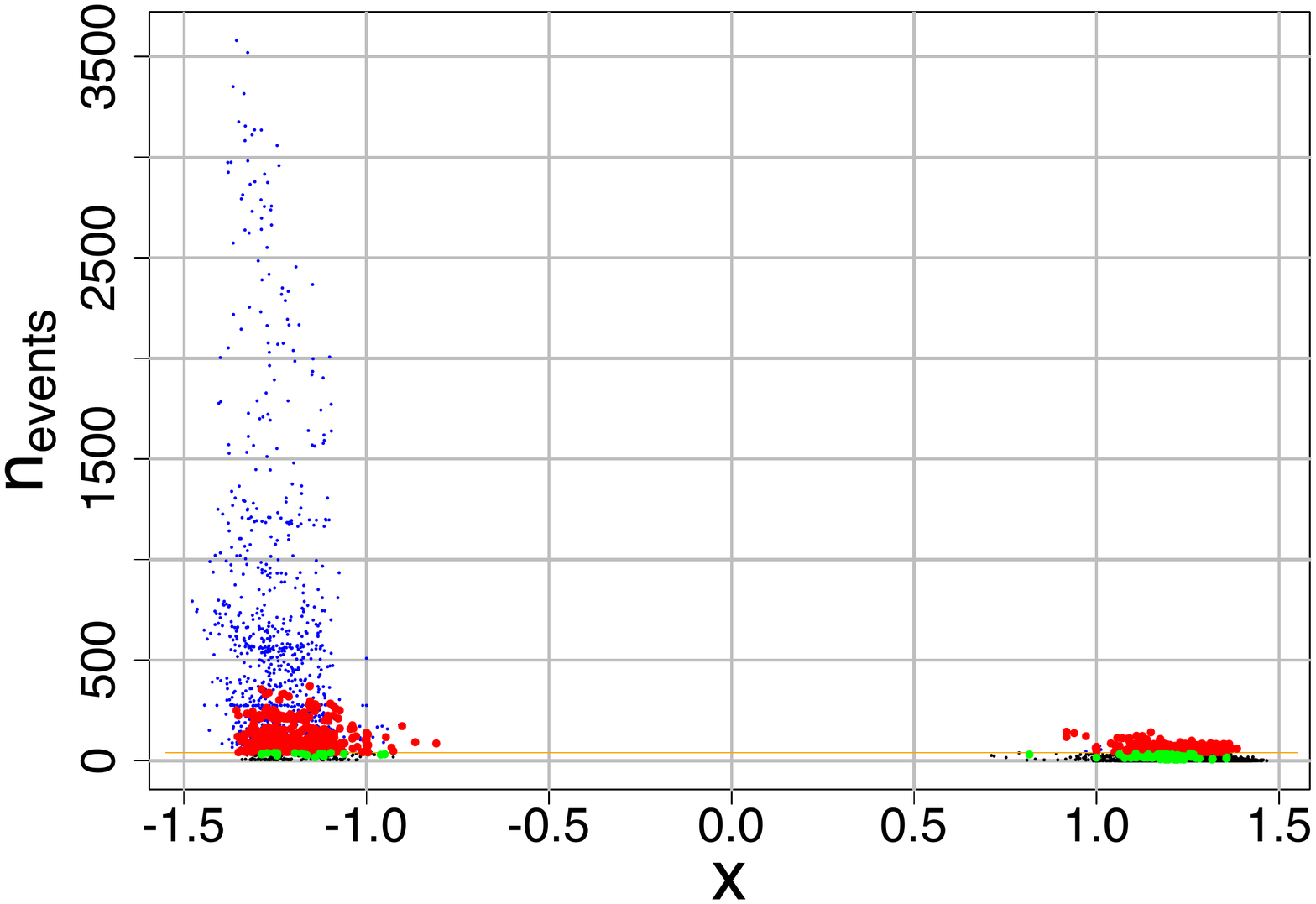}
     \includegraphics[width=6.8cm]{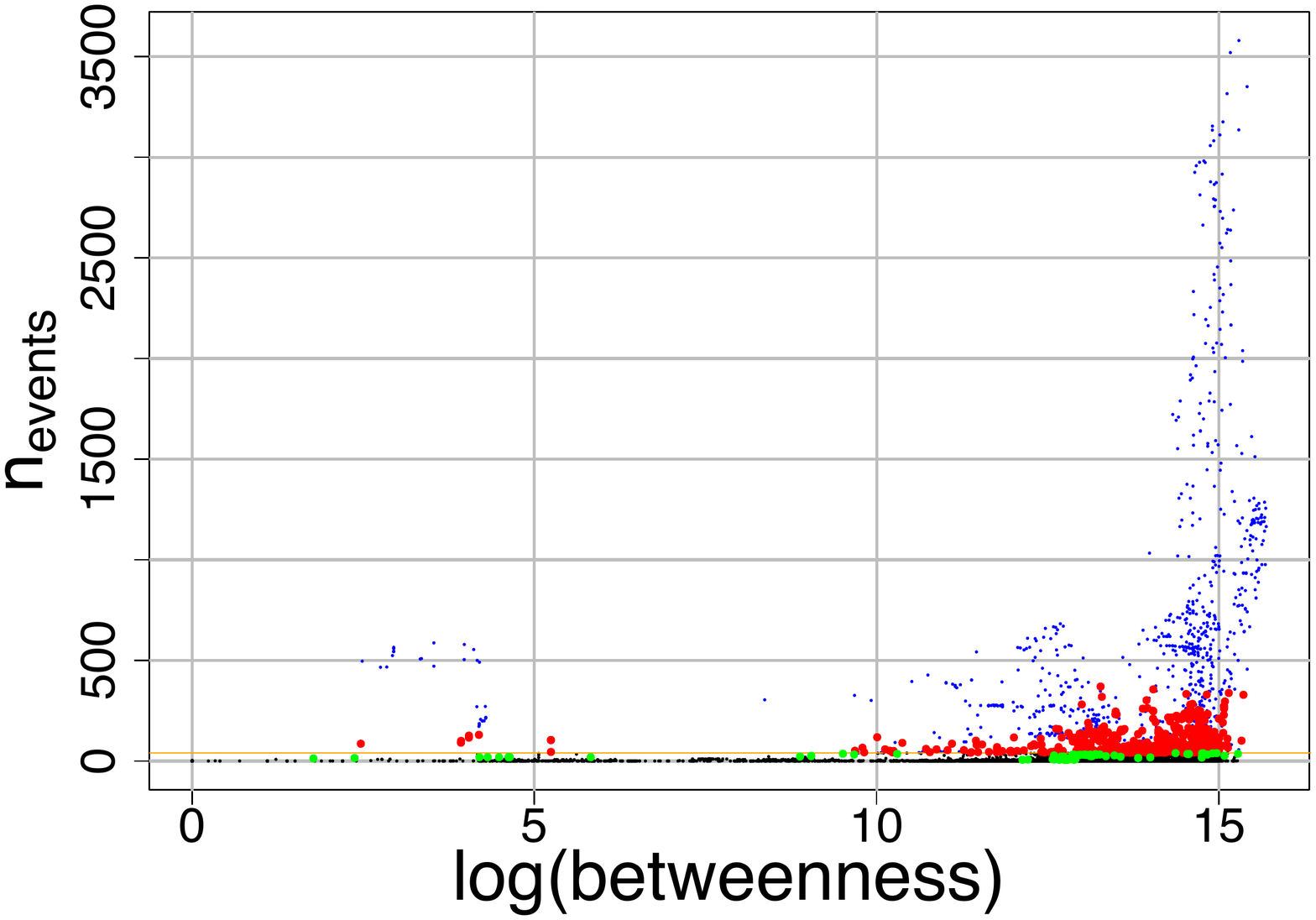}
  \end{subfigure}
\caption{Simulation for year $2015$ run after training over years $2002-2014$. (Left)Node coordinates , betweenness and number of conflict events are reported respectively on $x$,$y$ and $z$ axis. Bigger red and green dots indicate areas which have shifted respectively to conflict or peace state as compared to previous year.  Small blue and black dots indicate respectively ares of new conflict and peace.(Right) (Right) 2D projection on x-z plane (above) and z-y plane (below) of same plot. Orange line marks threshold on number of events.  (see Table  \ref{ttab1} , \ref{tab2}  for a quantitative analysis).}
\label{fig5}
\end{figure}



\subsection{Comparison with Linear Regression}
We  run  linear regression to see how  it compares to results of our model. 
We consider the number of conflict events  of e.g.  2015 and linearly fit to the number of events of previous year 2014. We then use
fitting parameters to predict events of  year 2016 based on previous year events.
In tables III and IV  we show how Linear Regression fares as compared to our model. Both in overall prediction and 
 on predicting  new events our model fares better (see tables \ref{ttab1},\ref{tab2}, \ref{tabbo}, \ref{tabbo2} ). In fact, on average per year our model fares better than Linear Regression in  terms of accuracy  (0.96 vs 0.95), of average F1 score (0.84 vs 0.81) and also on new events  (55 \%  vs 49\%), see also Fig. \ref{figBOX} for summary plot.Figures \ref{fig9} and \ref{fig10} (see Appendix) show examples of time series for model and LR for two cities.

\begin{table}[ht]
\centering
\begin{tabular}{rllllllllllllll}
  \hline
 & measure/year &  2005 & 2006 & 2007 & 2008 & 2009 & 2010 & 2011 & 2012 & 2013 & 2014 & 2015 & 2016 \\ 
  \hline
1 & Accuracy & 0.0.96 & 0.98 & 0.98 & 0.94 & 0.96 & 0.94 & 0.95 & 0.95 & 0.96 & 0.92 & 0.92 & 0.93 \\ 
  2 & F1 & 0 0.7 & 0.9 & 0.87 & 0.75 & 0.85 & 0.76 & 0.8 & 0.85 & 0.88 & 0.78 & 0.81 & 0.84 \\ 
  3 & Recall &  0.54 & 0.83 & 0.8 & 0.61 & 0.86 & 0.71 & 0.89 & 0.83 & 0.84 & 0.68 & 0.73 & 0.78 \\ 
  4 & Precision & 0.99 & 0.97 & 0.95 & 0.97 & 0.84 & 0.81 & 0.72 & 0.87 & 0.93 & 0.91 & 0.9 & 0.9 \\ 
   \hline
\end{tabular}
\caption{Accuracy, F1 score, recall and precision for predictions based on Linear Regression. Threshold for war/peace discrimination is 40 events}
\label{tabbo}
\end{table}


\begin{table}[h!]
\begin{tabular}{ |p{3cm}||p{1cm}|p{1cm}|p{1cm}|p{1cm}|p{1cm}| p{1.cm}|p{1cm}|p{1cm}|p{1cm}|p{1cm}|p{1cm}| p{1cm}| }
 \hline
 \multicolumn{13}{|c|}{Linear Regression performance on new conflict/peace occurrences} \\
 \hline
  year & 2005 & 2006 & 2007 & 2008 & 2009 & 2010 & 2011 & 2012 & 2013 & 2014 & 2015 & 2016 \\ 
 \hline
\% new  peace correct   & 100\%&  100\% &95\% &33\% & 12\%&4\% & 9\% &6\% &20\%& 44\% &64\%&48\%\\ 
\% new war correct & 56 \%&  61\%& 45\% &26\% &63\% & 36\% &62\% &61\% & 61\% &33\% &42\% &43\%\\
\% total new correct   & 69 \%&  73\%&59\% & 26\% &55\% &30\% & 44\% &53\% &56\% & 34\%& 43\% &44\% \\ 

 \hline
\end{tabular}
\caption{Performance of linear regression on predicting occurrence of new changes of node state to conflict or peace for years 2005-2016.}
\label{tabbo2}
\end{table}

\begin{figure}[h!]
  \includegraphics[width=6.8cm]{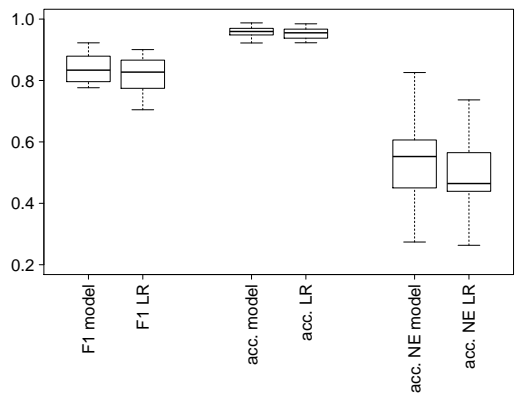}
\caption{Comparison of F1 score, total accuracy and  accuracy on predicting new events (NE) between model and Linear Regression (LR).}
\label{figBOX}
\end{figure}

    \section{Betweenness individuates high conflict areas in three-state classification}
Betweenness or betweenness centrality is a quantity defined for each node of a network as the sum of all the shortest paths between two other nodes of the network passing through the node in consideration.
More precisely the betweenness centrality $b(i)$ of node $i$ is defined as
\begin{equation}
    b(i)=\sum_{j\neq i \neq k}\frac{ \sigma_{jk}(i)}{\sigma_{jk}}
\end{equation}
where  $\sigma_{jk}$ is the total number of shortest paths between nodes $j$ and $k$ and  $\sigma_{jk}(i)$  is the number of  shortest paths between $j$ and $k$ that pass through node $i$. 
As shown in \cite{Guo17} it is apparent that conflict areas with higher number of conflict events tend to coincide with
network nodes with higher betweenness.
Here we aim at quantifying this trend by introducing a third classification state, high intensity conflict , and using betweenness as an indicator of this state.
So we classify as peaceful areas  those nodes  with $x>0$ and then we identify as low conflict areas those with $x<0$  and   betweenness$<$threshold,  and high intensity conflict areas, those
with $x<0$ AND betweenness$>$threshold. The thresholds adopted are $40$ events  per year for the low conflict threshold and $770$ events per year for the high conflict data.
We determine the latter as average over the threshold values that maximise the $F1$ score  for the predictions for each year.

\begin{figure}[h!]

\begin{subfigure}{7.8cm}
\hspace{-7.cm}  \includegraphics[width=14cm]{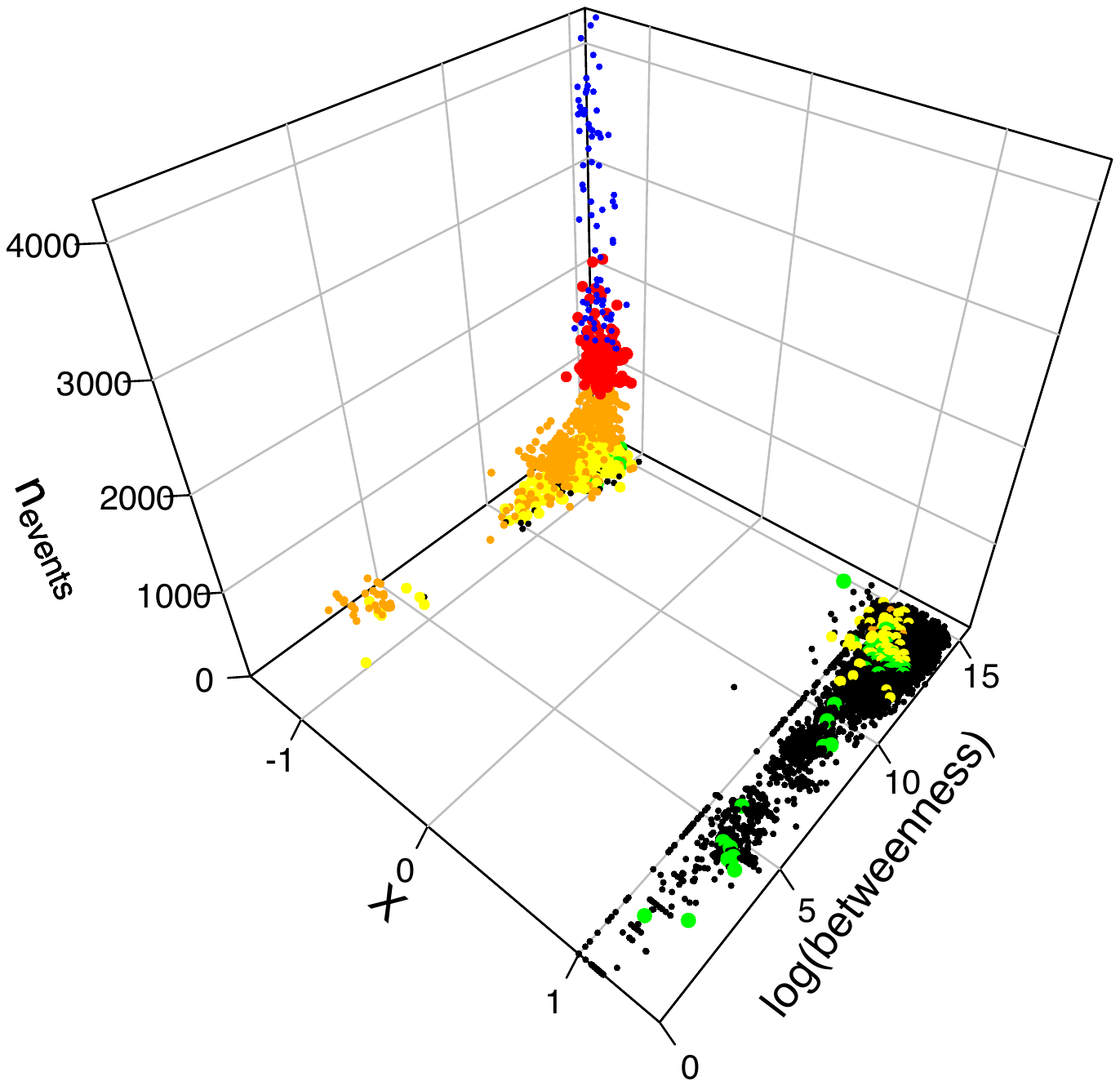}
  \end{subfigure}
  \begin{subfigure}{5.2 cm}
    \includegraphics[width=6.8cm]{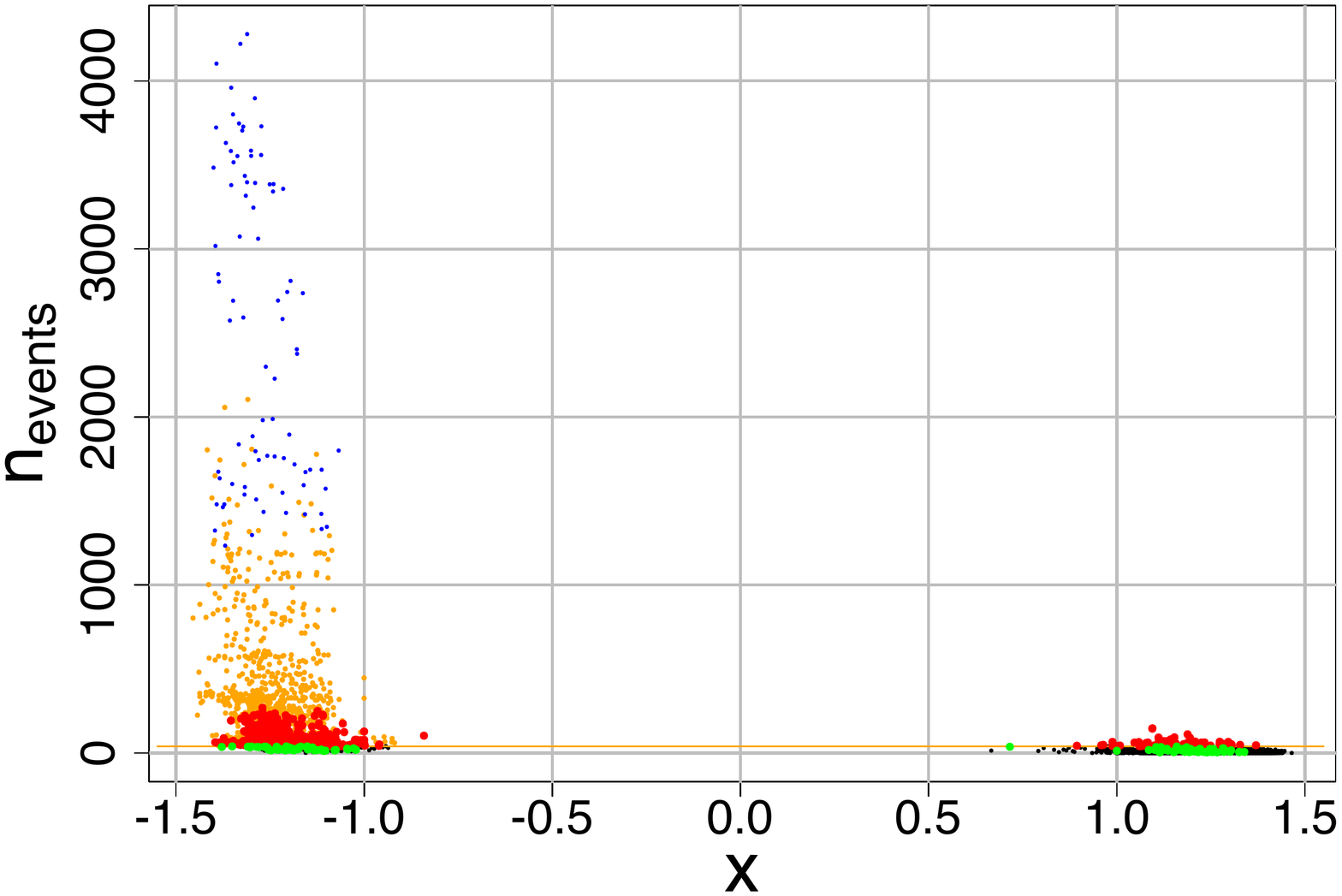}
     \includegraphics[width=6.8cm]{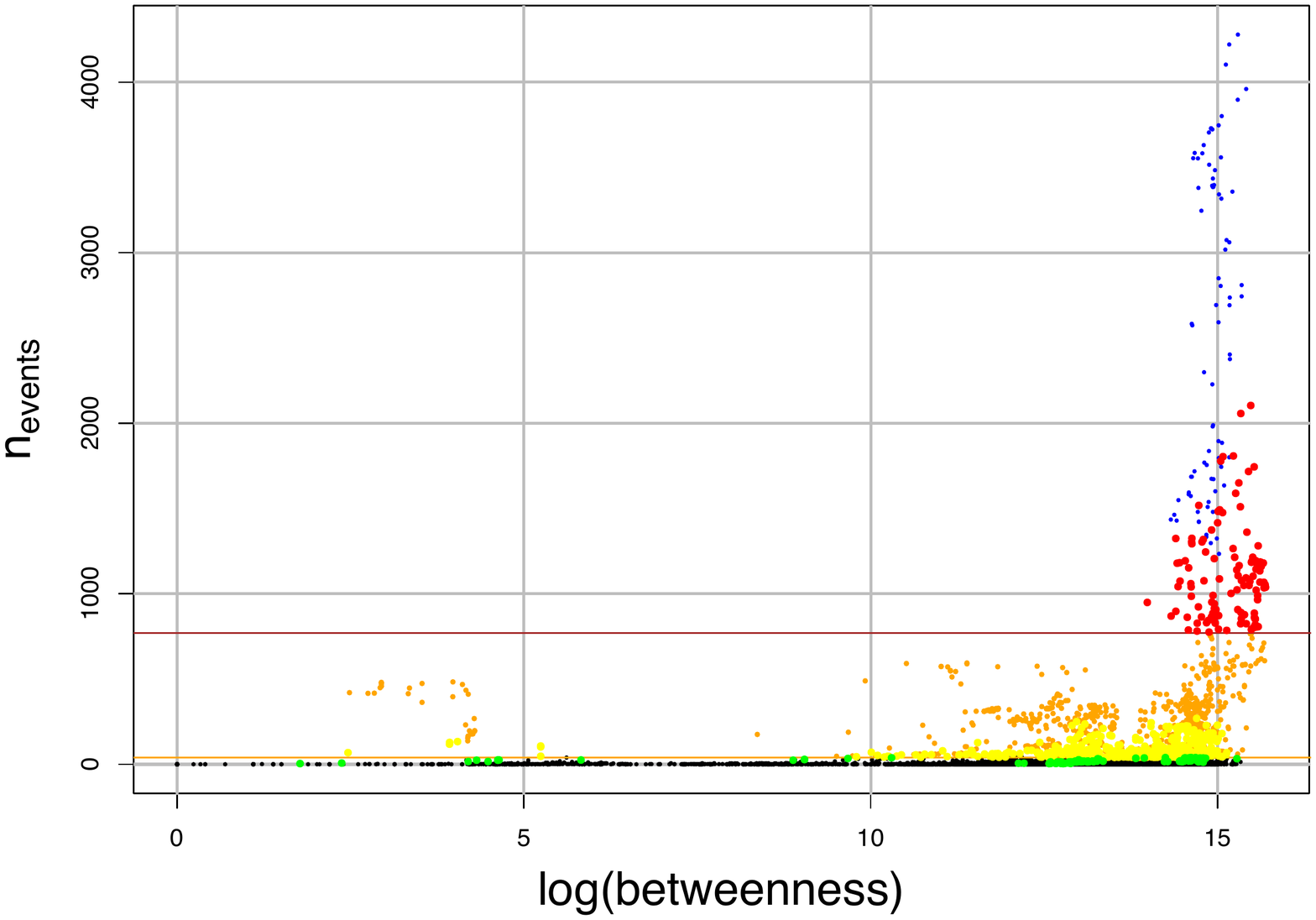}
  \end{subfigure}
\caption{Simulation for year $2016$ run after training over years $2002-2016$. Node coordinates , betweenness and number of conflict events are reported respectively on $x$,$y$ and $z$ axis. Bigger red, yellow and green dots indicate areas which have shifted respectively to conflict or peace state as compared to previous years.  Small blue and black dots indicate respectively ares of conflict and peace, which have not changed  state as compared to previous year.  (Right) 2D projection on x-z plane of same plot. Orange line marks threshold on number of events.  (see Table V for a quantitative analysis.}
\label{fig6}
\end{figure}

By defining the relative  $3\times3$ confusion matrix $M$  which in element $M_{ij}$ has the number of actual events of type $j$ classified as $i$. 
One can then evaluate true and false positives and true and false negatives and consequently precision, recall and F1 score.
A measure of how the model performs is therefore obtained.

We find an improved overall  $F1$ score.
Table \ref{TABB} show F1 score for this non-binary classification and performance on new conflict and peace events on the predictions for year 2005-2016. 
  Figure \ref{fig6} shows how the colouring of Figure \ref{fig4} obtained for binary classification, changes when taking into account betweennes and high-conflict areas in a three-state classification.
From  Table \ref{TABB} it is also apparent how  the accuracy of the model in predicting new high conflict states becomes high when using betweenness as classifier.
This strengthens the importance of betweenness in individuating high conflict areas.

\begin{table}[h!]
\begin{tabular}{rlllllllllllll}

  \hline
 & measure/year & 2005 & 2006 & 2007 & 2008 & 2009 & 2010 & 2011 & 2012 & 2013 & 2014 & 2015 & 2016 \\ 
  \hline
1 & Accuracy & 0.98 & 0.98 & 0.98 & 0.98 & 0.97 & 0.96 & 0.95 & 0.95 & 0.96 & 0.95 & 0.95 & 0.96 \\ 
  2 & F1 & 0.95 & 0.96 & 0.97 & 0.97 & 0.94 & 0.94 & 0.94 & 0.93 & 0.94 & 0.94 & 0.91 & 0.94 \\ 
     \hline
\end{tabular}
\caption{Accuracy and F1 score  for predictions based on model after training over previous years. Threshold for peace, low and high conflict discrimination are 40  and 770 events respectively.}
\label{TABB}
\end{table}

\begin{figure}[h!]
\includegraphics[height=9 cm,width=11.2 cm, angle=0]{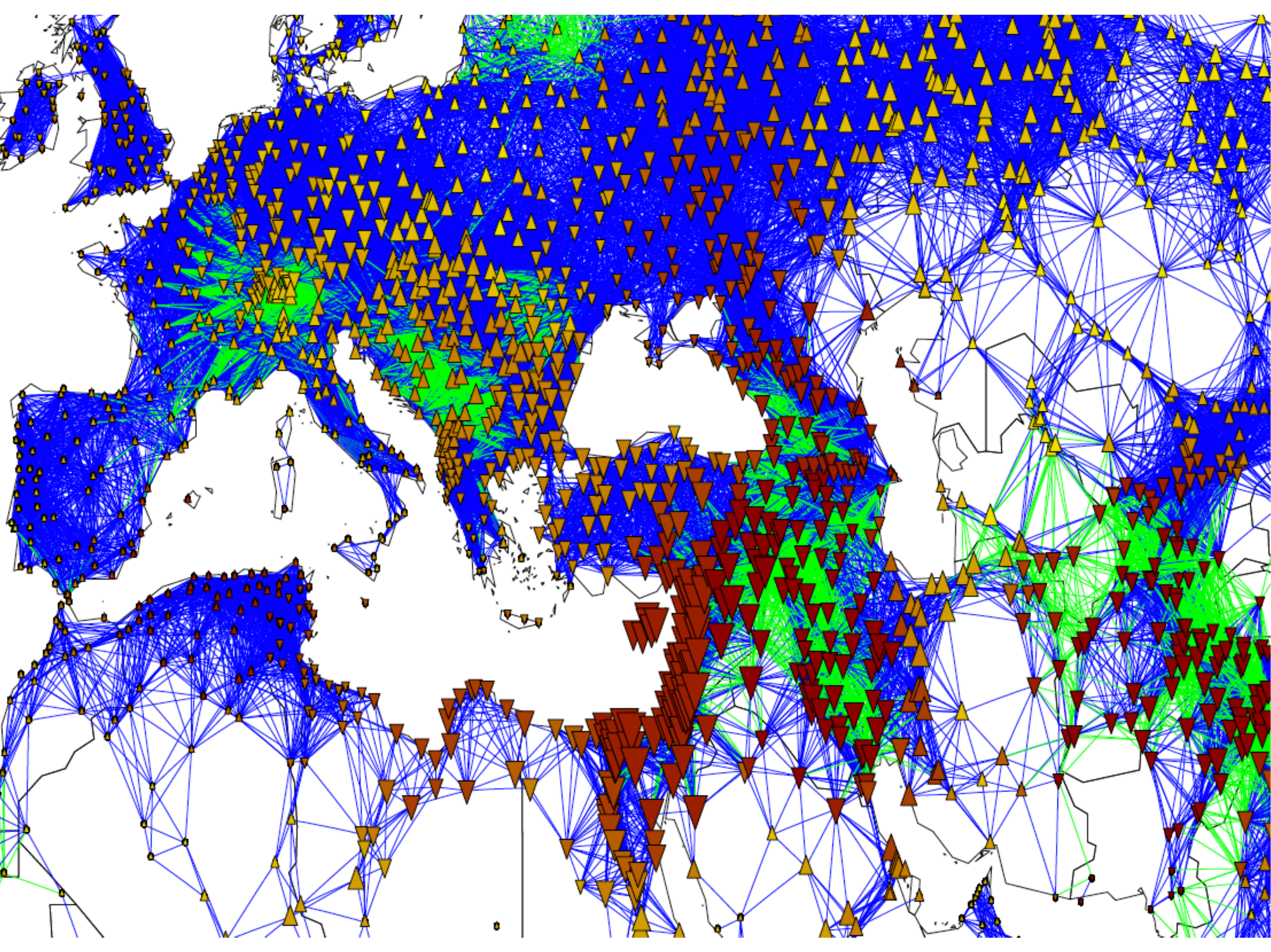}
\caption{Simulation for year $2016$ run after training over years $2002-2015$ mapped on the earth surface. Nodes are represented as triangles. Colour of nodes goes from yellow to red with increasing number of conflict events. Size of  triangles is proportional to the logarithm of betweenness. Triangles pointing down indicate conflict area as resulting from model prediction. Triangles pointing up indicate peace area. Blu links are alliances while green links are non-collaborative country relations.}
\label{fig7}
\end{figure}

In Figure \ref{fig7} we plot the results of simulation of Figure 4 mapped on the earth surface, this figure gives a visual representation of how nodes with high betweenness strongly correlate with  high conflict, as resulting from quantitative analysis.

\section{Tracing causal links to new war and peace events}
The interesting feature of our dynamical model  is that allows us to trace back the transition of a node state to its dynamical
cause. We achieve this by considering a node that changes state as compared to previous year  and disconnecting the neighbour nodes
in a systematic way to see how this changes the final steady state of the node of interest. 
More specifically we disconnect all neighbours from same country at the same time and run simulation and observe if a change in the final state of the node occurs. 
\begin{table}[h!]
\label{somalia}
	\begin{minipage}{0.8\linewidth}
		
		\centering
		  \begin{tabular}{ |p{0.8cm}||p{1.5cm}|p{1.5cm}|p{2.5cm}|p{2.3cm}|p{2cm}|p{2cm}|   }
 \hline
 \multicolumn{7}{|c|}{ Causal links to new events, Somalia. Overall accuracy on new events: 65\%} \\
  \hline
  Year & city & current state & previous year state  & layer disconnected & country disconnected & current state with layer disconnected \\ 
  \hline
 2008 & Buur Gaabo & War &Peace& Geographic & Somalia & Peace \\ 
 2008 & Buur Gaabo & War &Peace & Political & Kenia (E)  & Peace \\ 
 2013 & Hurdiyo & War&Peace & Geographical & Somalia & Peace \\ 
 2013 & Hurdiyo & War &Peace& Geographical & Somaliland(E) & Peace \\ 
 2009 & Gaalkacyo & War &Peace& Geographical & Ethiopia (E)+ Somalia+ Somaliland (E) & Peace \\ 
 2009 & Eyl & War&Peace & Geographical & Somalia+ Somaliland & Peace \\ 
   \hline
\end{tabular}
	\end{minipage}\hfill
	\begin{minipage}{0.2\linewidth}
		\centering
\includegraphics[height=9 cm,width=6 cm,angle=0]{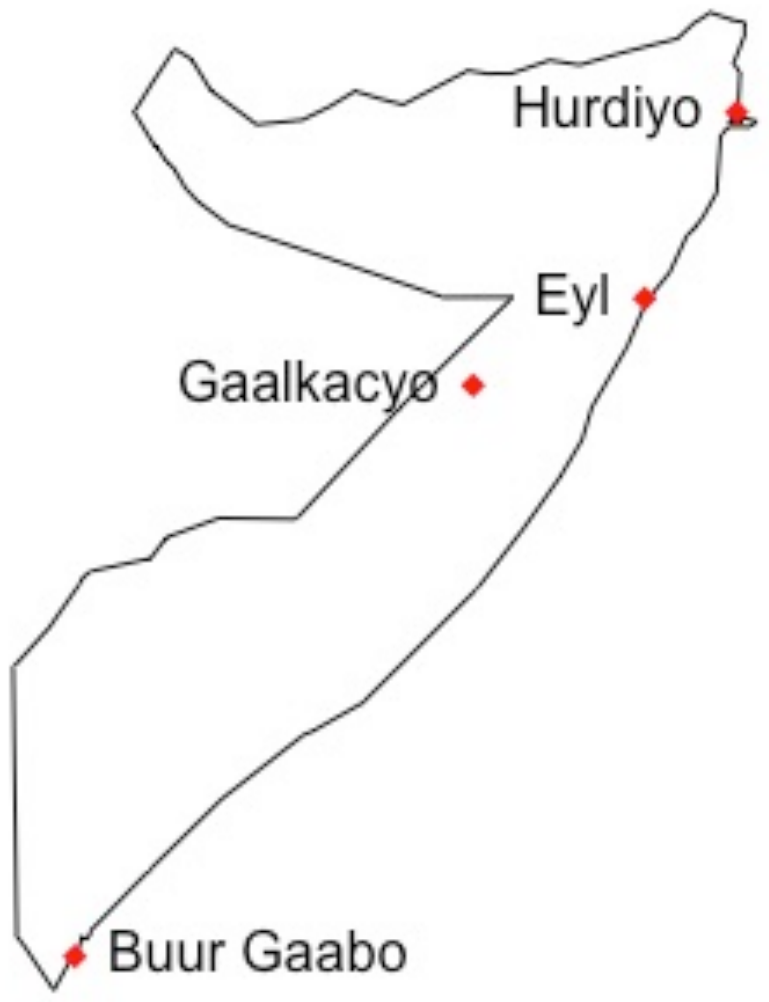}
	\end{minipage}
		\vspace{-3cm}
		
			\begin{minipage}{0.66\linewidth}

				 \begin{tabular}{ |p{0.8cm}||p{1.5cm}|p{1.5cm}|p{2.5cm}|p{2.3cm}|p{2cm}|p{2cm}| }
 \hline
 \multicolumn{7}{|c|}{ Causal links to new events, Myamnar.Overall accuracy on new events: 49\%} \\
  \hline
 Year & city & current state & previous  state  & layer disconnected & country disconnected & current state with layer disconnected \\ 
     \hline
    2013 & Sittwe & War & Peace & Geographic & Bangladesh (E) +India(E) +Myanmar & Peace \\ 
  2014 & Magway & War & Peace & Geographic & Myanmar & Peace \\ 
  2015 & Kyaukphyu & War & Peace & Geographic & Myanmar & Peace \\ 
  2013 & Taunggyi & War & Peace & Geographic & Myanmar & Peace \\ 
   \hline
\end{tabular}
\end{minipage}\hfill
		\begin{minipage}{0.33\linewidth}
		\centering\includegraphics[height=10 cm,width=10cm,angle=0]{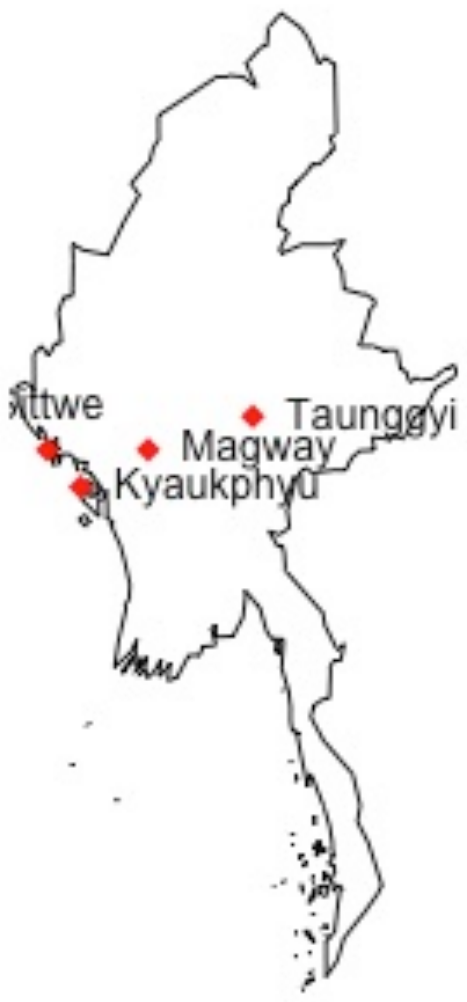}
				
					\end{minipage}
	
		\begin{minipage}{0.64\linewidth}
	
			\vspace{-4cm}			
 \begin{tabular}{ |p{0.8cm}||p{1.5cm}|p{1.5cm}|p{2.5cm}|p{2.3cm}|p{2cm}|p{2cm}| }
 \hline
 \multicolumn{7}{|c|}{ Causal links to new events: Colombia. Overall accuracy on new events: 65\%} \\
  \hline
Year & city & current state & previous year state  & layer disconnected & country disconnected & current state with layer disconnected \\ 
  \hline
2014 & Jurado & War & Peace & Geographic & Colombia & Peace \\ 
  2011 & Valledupar & Peace & War & Political & Colombia & War \\ 
  2011 & Valledupar & Peace & War & Political & Venezuela (F) & War \\ 
  2012 & Tame & War & Peace & Geographic & Colombia & Peace \\ 
  2012 & Orocue & War & Peace & Geographic & Colombia+  Venezuela (F) & Peace \\ 
  2014 & Santa Marta & War & Peace & Geographic & Colombia & Peace \\ 
   \hline
\end{tabular}
	\caption{}
\end{minipage}\hfill
	\begin{minipage}{0.32\linewidth}
\vspace{-4cm}
			\centering\includegraphics[height=11 cm,width=9cm,angle=0]{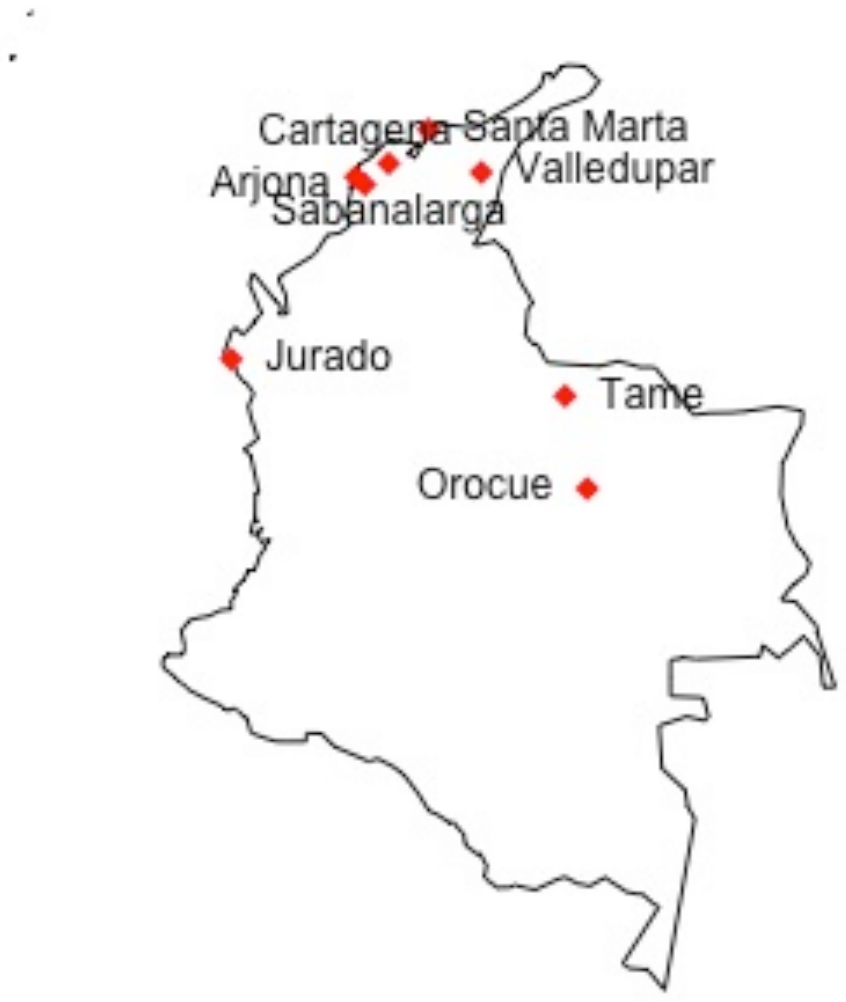}
		
	\end{minipage}
		\end{table}

We repeat the same operation for  the  the geographical, political and cultural layers to see if a change occurs. For the geographical level links are removed, for the political layer
the alliance is shifted, while for the cultural layer  no difference between populations is assumed.
This allows us to trace back the cause of the change of state of the node. 
We focus here on five countries and illustrate the results stemming from our model.

In the case of Somalia , for example,  where our model is able to predict new events with an overall accuracy of  $95\%$ we see that for the city of Buur Gaabo, disconnecting at the political level, i.e. shifting relationship of Kenia from enemy (E) to ally (F)
induce a change in the node state from war to peace. This  linking a possible tensions at the border as leading to conflict.  The fact that disconnecting the geographic layer for the same
city with neighbour nodes of the same country, leads to  a peace state, on the other hand, points towards more the long-standing internal struggles of Somalia.

In the case of Myanmar on the other hand we see that only disconnecting the geographic layer leads to a shift  of outcome in the steady states of the nodes.
In all cases illustrated  disconnecting neighbours of same countries leads to peace, hinting again at internal causes of conflict. Only in the case of Sittwe
we see that to induce change disconnecting links to Bangladesh and India, two enemy countries, is necessary. Finally in Colombia we see that  the shift of alliance of Venezuela causes change from peace to war state in Valledupar in 2011, while in all other cases  disconnecting links to neighbours of same country changes node state from war to peace, hinting at internal source of conflict.

\section{Conclusions}
In this paper have introduced a new model for predicting conflict. Starting from a geospatial network of cities  and adding  political and cultural layers we have implemented on a multiplex network structure  a dynamical model describing the evolution of the state of each city/node  between a conflict or peace state due to interaction with other nodes. We have taken into account political alliances and cultural differences between populations  and  shown that our model is able to predict with high accuracy the well-being state of each node.  
The model  overall performs  better than linear regression   and performs rather well on predicting  new occurrences of peace and war, which is less trivial.
Adding betweenness as a classifier of high conflict states increases  the overall performance of the model induced classification, confirming the observed link between  high conflict areas and betweenness centrality as first observed in \cite{Guo17}. Importantly our model allows one to trace back in many cases the causal link to the occurrence of change of state in a node, which is crucial as a more advanced version of the model could allow one to develop  a better understanding of the dynamics leading to conflicts in risk areas, informing more effective preventive policies.
Our model offer a promising platform for predicting conflict and can be further improved with additional feature and layers, e.g. using data of economic growth or resource availability for the field  term $H$, or adding maritime trade data.
This will lead to further improved performance and are part of future development.

\appendix*
\section{Libya and Yemen}
Here are reported the results  obtained  running simulations and disconnecting layers in nodes where there is a change of state,  for the states of Libya and Yemen.
In both cases main causal link for change of state is traced in the interaction with  cities from the same country, hinting at internal source of conflict.
\begin{table}[h!]
	\begin{minipage}{0.7\linewidth}
				 \begin{tabular}{ |p{0.8cm}||p{1.5cm}|p{1.5cm}|p{2.5cm}|p{2.3cm}|p{2cm}|p{2cm}| }

 \hline
 \multicolumn{7}{|c|}{ Causal links to new events, Libya. Overall accuracy on new events: 56\%} \\
\hline
Year & city & current state & previous year state  & layer disconnected & country disconnected & current state with layer disconnected \\ 
  \hline
2016 & Tmassah & War & Peace & Geographic & Libya & Peace \\ 
  2014 & Hun & War & Peace & Geographic & Libya & Peace \\ 
    2014 & Ghadamis & War & Peace & any & any & No change \\ 
  2014 & Hun & War & Peace & any & any & No change \\   
   \hline
\end{tabular}
\end{minipage}\hfill
	\begin{minipage}{0.3\linewidth}

		\centering\includegraphics[height=9 cm,width=8cm,angle=0]{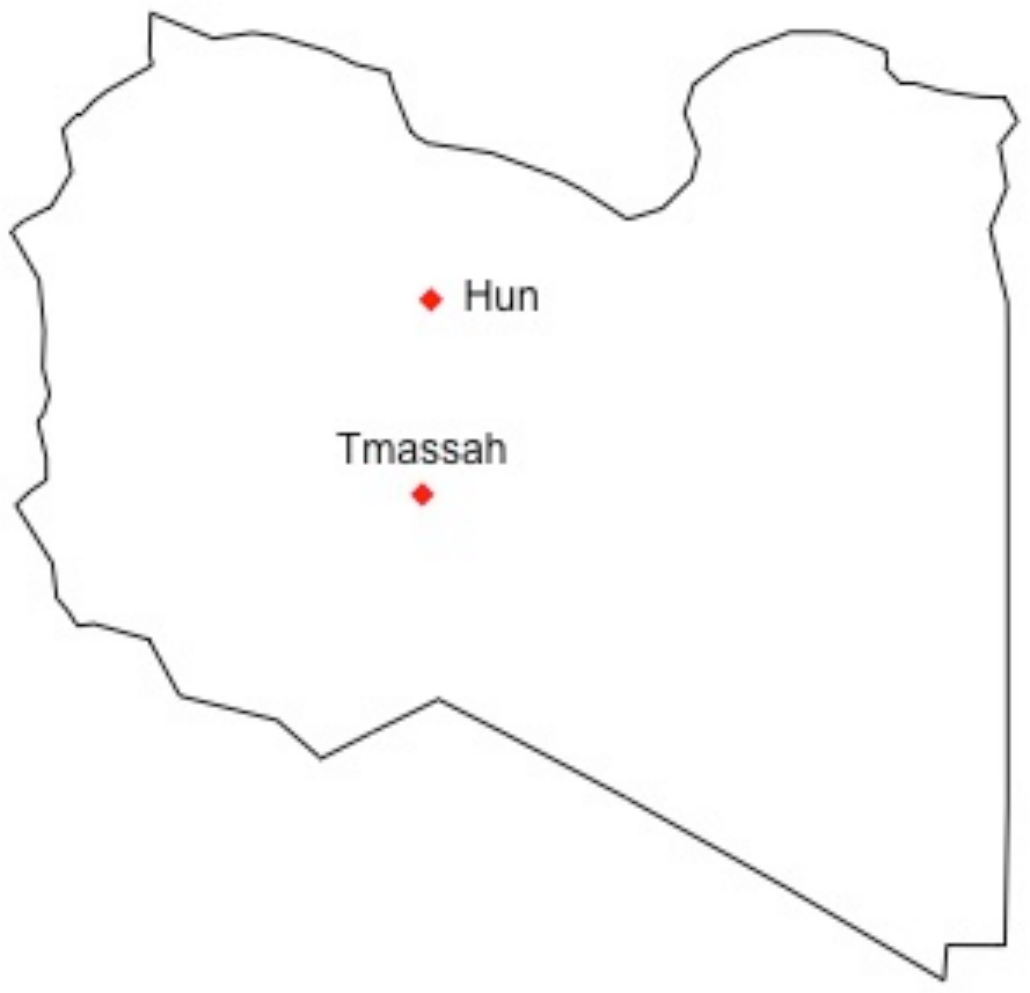}
				\vspace{-5cm}
			\end{minipage}
			
	\begin{minipage}{0.65\linewidth}
				 \begin{tabular}{ |p{0.8cm}||p{1.75cm}|p{1.5cm}|p{2.5cm}|p{2.3cm}|p{2cm}|p{2cm}| }
 \hline
 \multicolumn{7}{|c|}{ Causal links to new events, Yemen. Overall accuracy on new events: 71\%} \\
 \hline
Year & city & current state & previous year state  & layer disconnected & country disconnected & current state with layer disconnected \\ 
  \hline
2013 & Al Ghaydah & War & Peace & Geographic & Yemen & Peace \\ 
  2011 & Al Hudaydah & War & Peace & any & any & No change \\ 
  2011 & Sadah & War & Peace & any & any & No change \\ 
  2011 & Ibb & War & Peace & any & any & No change \\ 
   \hline
\end{tabular}
\caption{}
\end{minipage}\hfill
		\begin{minipage}{0.25\linewidth}
		\centering\includegraphics[height=8cm,width=6cm,angle=0]{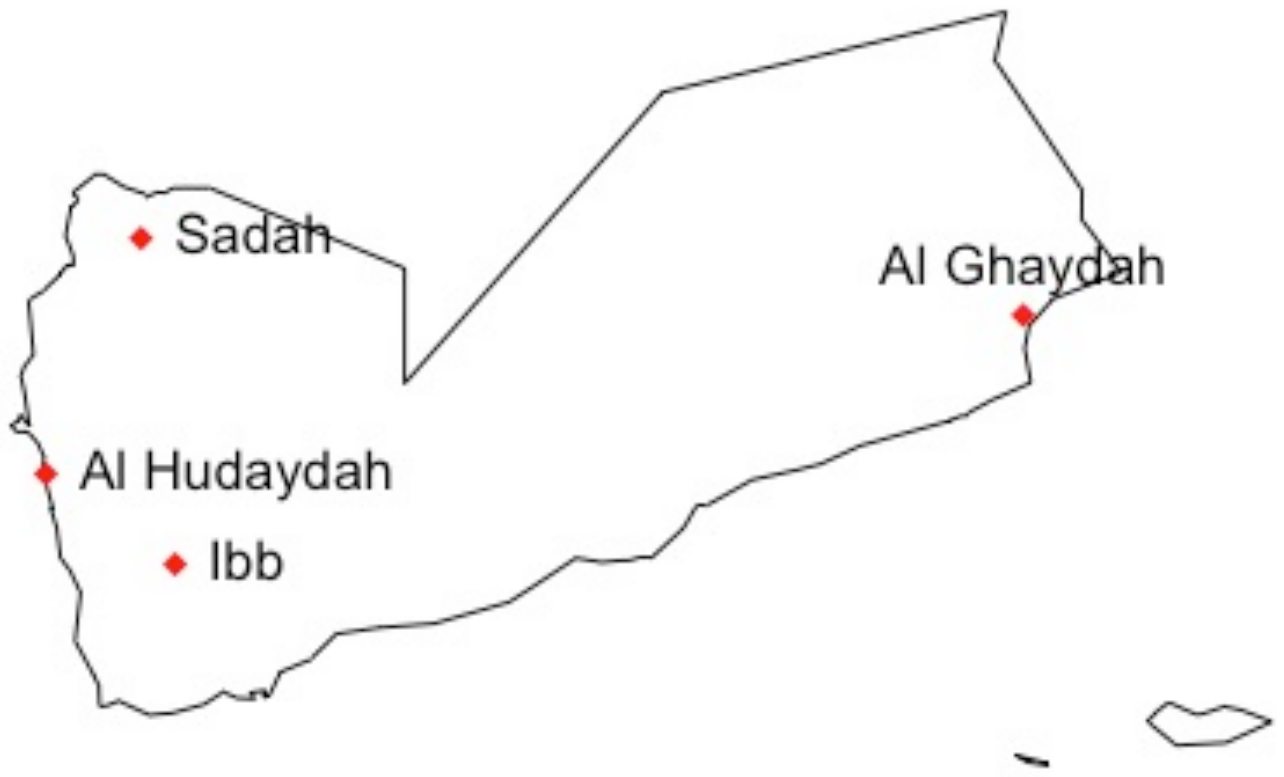}
		\label{ }
	\end{minipage}

	\end{table}
\vspace{9cm}

\begin{figure}[h!]

\begin{subfigure}{5.8cm}
 \hspace{-6.7cm}    \includegraphics[width=7.7cm]{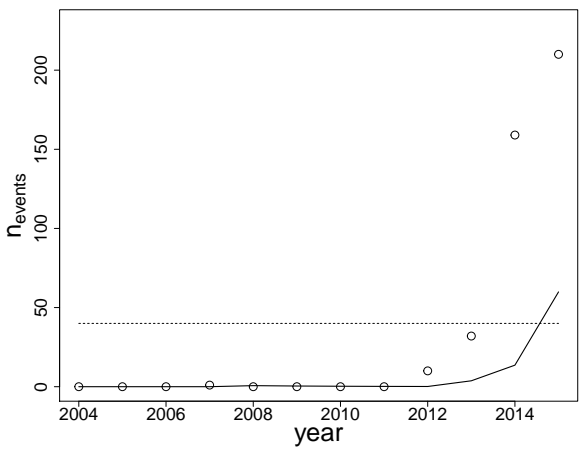}
  \end{subfigure}
  \begin{subfigure}{5.0 cm}
    \includegraphics[width=7.7cm]{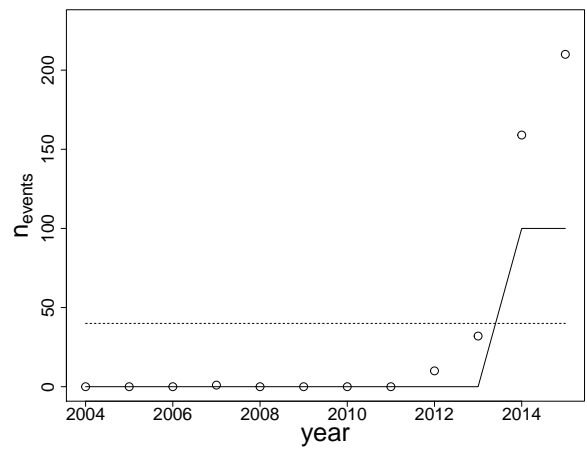}
  \end{subfigure}
\caption{Time series and results for Linear Regression (left) and model (right) for the city of Hun, Libya. To compare with data, model output is here set to 0 for peace state and 100 for war state. Dashed line shows threshold on number of events. }
 \label{fig9}
\end{figure}
\begin{figure}[h!]
\begin{subfigure}{5.8cm}
 \hspace{-6.7cm}    \includegraphics[width=7.7cm]{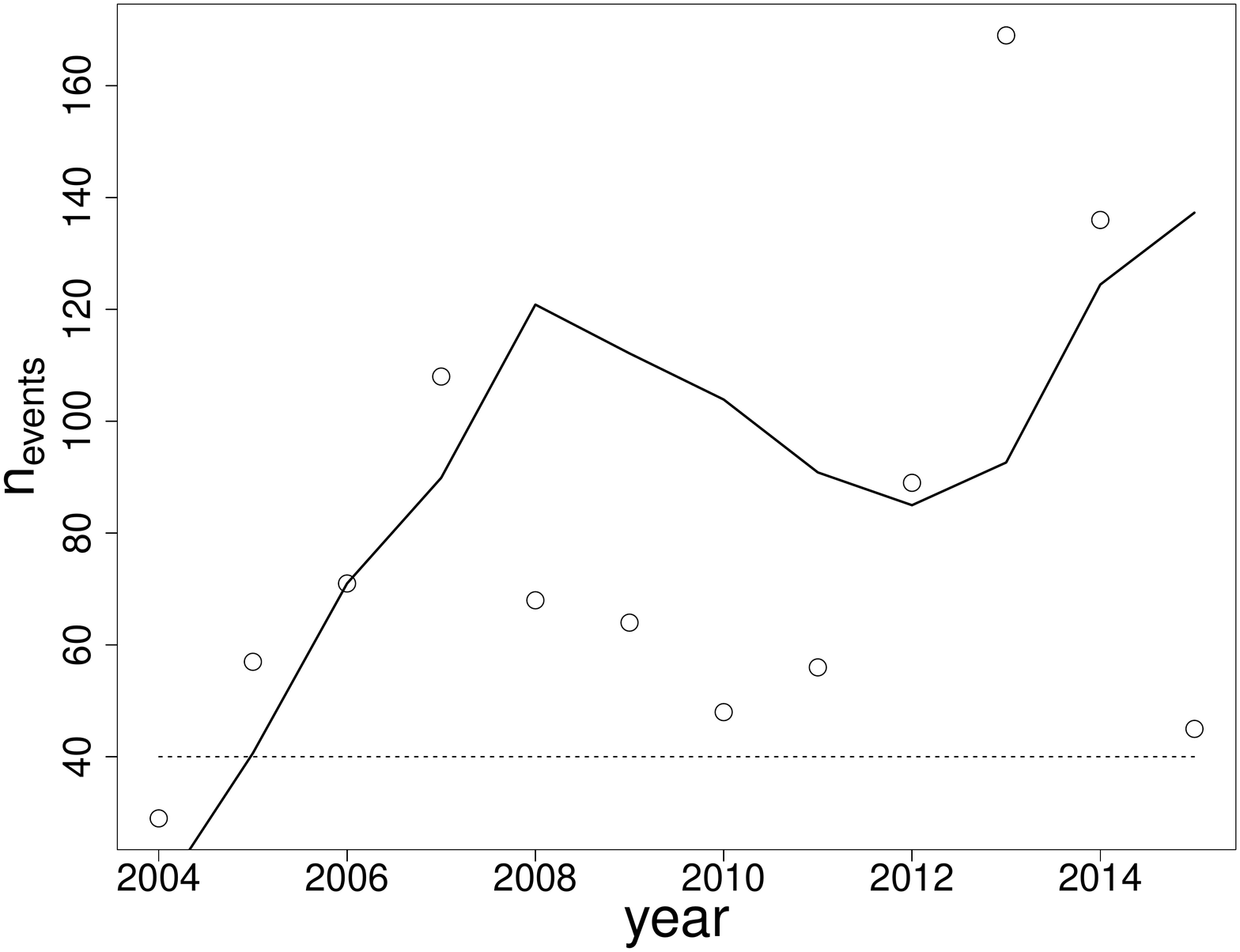}
  \end{subfigure}
  \begin{subfigure}{5.0 cm}
    \includegraphics[width=7.7cm]{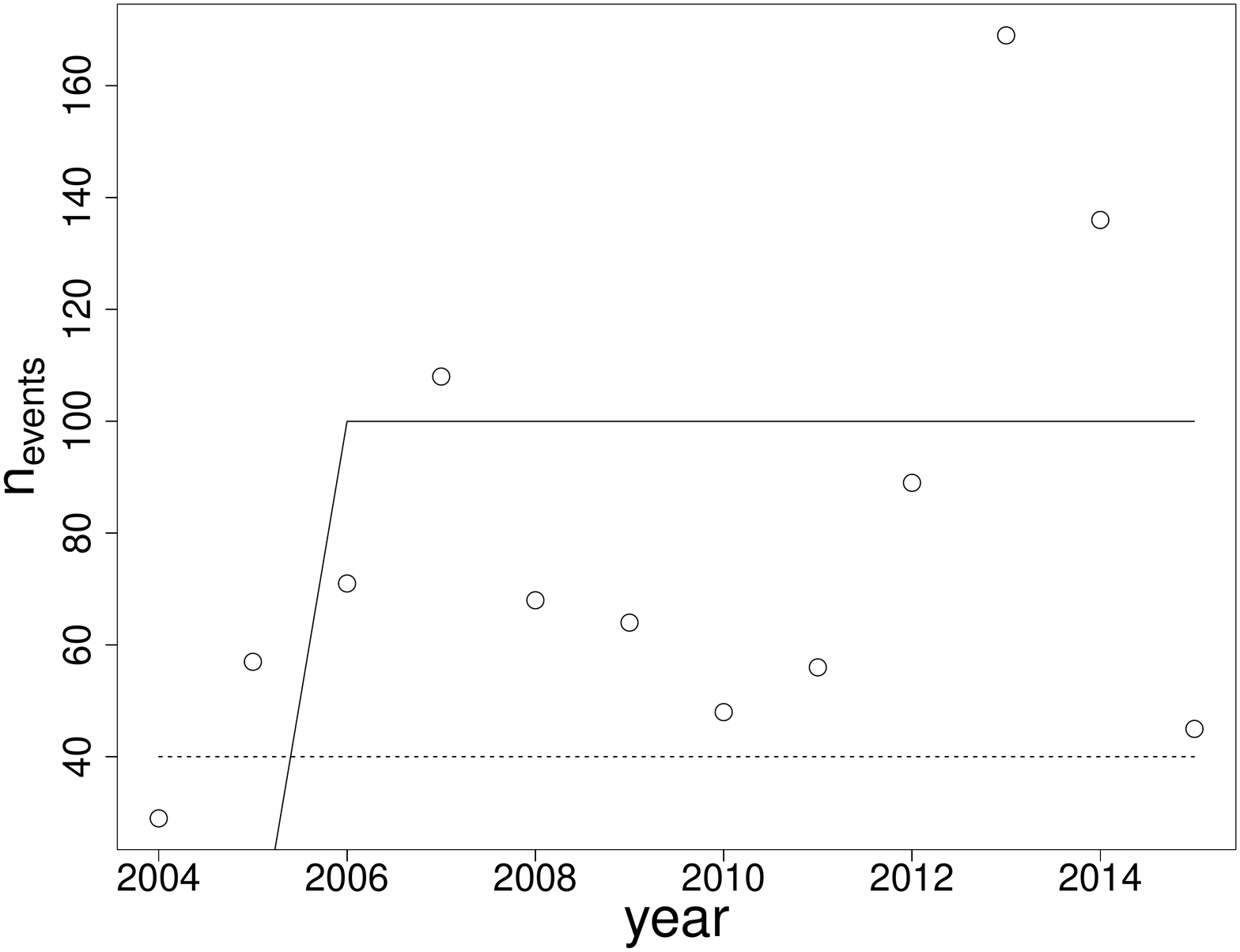}
  \end{subfigure}

\caption{Same as Fig 8: Time series and results for Linear Regression (left) and model (right) for the city of Al Jahra,Kuwait}
  \label{fig10}
\end{figure}

\end{document}